\documentclass[journal]{IEEEtran}

\usepackage[numbers]{natbib}
\usepackage[utf8]{inputenc}
\usepackage{tikz}
\usepackage{pgfplots}
\usepackage{tikz-3dplot}
\usepackage{textcomp}
\usepackage{lineno}
\usepackage{booktabs}
\usepackage{balance} 
\usepackage{tabularx}
\usepackage{mathtools}
\usepackage{graphicx}
\usepackage{graphics}
\usepackage{epstopdf}
\usepackage{amsmath,amssymb,amsfonts}
\usepackage[english]{babel}
\usepackage{amsthm}
\usepackage{soul}
\usepackage{color}
\usepackage{subcaption}
\usepackage{wrapfig}
\usepackage{comment}
\usepackage{float}
\usepackage{amssymb}
\usepackage{xcolor}
\usepackage{lipsum}
\usepackage{enumitem}
\usepackage{bbm}
\usepackage{hyperref}
\usepackage{pdflscape}
\usepackage{natbib} 
\usepackage{algorithm}
\usepackage{algpseudocode}

\newcommand\undermat[2]{%
  \makebox[0pt][l]{$\smash{\underbrace{\phantom{%
    \begin{matrix}#2\end{matrix}}}_{\text{$#1$}}}$}#2}

\newtheorem{remark}{Remark}

\modulolinenumbers[5]
\usetikzlibrary{calc,patterns,angles,quotes}
\tdplotsetmaincoords{60}{115}
\pgfplotsset{compat=newest}

\begin{document}

\title{LoS-Map Construction for Proactive Relay of Opportunity Selection in 6G V2X Systems}

\author{Francesco Linsalata,~\IEEEmembership{Student Member,~IEEE,}
        Silvia Mura,~\IEEEmembership{Student Member,~IEEE,}\\~Marouan Mizmizi,~\IEEEmembership{Member,~IEEE,}~Maurizio Magarini,~\IEEEmembership{Member,~IEEE,}
        Peng Wang,~\IEEEmembership{Member,~IEEE,}~\\Majid Nasiri Khormuji,~\IEEEmembership{Senior Member,~IEEE,}~Alberto Perotti,~\IEEEmembership{Senior Member,~IEEE,}
        \\Umberto Spagnolini,~\IEEEmembership{Senior Member,~IEEE}
        
\thanks{F. Linsalata, S. Mura, M. Mizmizi,  and M. Magarini are with Dipartimento di Elettronica, Informazione e Bioingegneria, Politecnico di Milano, Via Ponzio 34/5, 20133, Milano
Italy.}
\thanks{P. Wang, M. N. Nasiri, and A. Perotti are with Huawei Technologies Sweden AB, Skalholtsgatan 9-11, SE-164 94 Kista, Stockholm, Sweden}
\thanks{U. Spagnolini is with Dipartimento di Elettronica, Informazione e Bioingegneria, Politecnico di Milano, Via Ponzio 34/5, 20133, Milano Italy and Huawei Industry Chair.}}

\maketitle

\begin{abstract}
Recent advances in Vehicle-to-Everything (V2X) technology and the upcoming sixth-generation (6G) network will dawn a new era for vehicular services with enhanced communication capabilities.
Connected and Autonomous Vehicles (CAVs) are expected to deliver a new transportation experience, increasing the safety and efficiency of road networks.
The use of millimeter-wave (mmW) frequencies guarantees a huge amount of bandwidth ($>$1GHz) and a high data rate ($>$ 10 Gbit/s), which are required for CAVs applications. 
However, high frequency is impaired by severe path loss, and line of sight (LoS) propagation can be easily blocked by static and dynamic obstacles.
Several solutions are being investigated, and the most promising one exploits relays. However, traditional relay schemes react to link failure and leverage instantaneous information, which impedes efficient relay selection in highly mobile and complex networks, such as vehicular scenarios. 
In this context, we propose a novel proactive relaying strategy that exploits the cooperation between CAVs and environment information to predict the dynamic \textit{LoS-map}, which describes the links' evolution in time. 
The proactive relaying schemes exploit the dynamic \textit{LoS-map} to maximize the network connectivity.
A novel framework integrating realistic mobility patterns and geometric channel propagation models is proposed to analyze the performance in different scenarios. 
Numerical simulations suggest that the proactive relaying schemes mitigate beam blockage and maximize the average probability of connecting CAVs with reliable links.
\end{abstract}


\IEEEpeerreviewmaketitle

\section{Introduction}

\IEEEPARstart{C}{onnected} and Automated Vehicle (CAV) technology will revolutionize the transportation system by providing a safer, more efficient, and less polluted road environment. CAV's main distinguishing features are the availability of various on-board sensors (e.g., Lidar, Radar, cameras, etc.) for environment perception and the enhancement of the Vehicle-to-Everything (V2X) communication capabilities~\cite{ Iwashina8300313, survey_aut_drive}. 
Current V2X systems are mainly based on two alternative radio access technologies: i) Dedicated Short Range Communications (DSRC) sponsored by IEEE~\cite{TVT_dsrc} and ii) Cellular V2X (C-V2X) promoted by the 3rd Generation Partnership Project (3GPP)~\cite{Sepulcre8691973}. Both standards operate at sub-6 GHz frequencies, i.e., Frequency Range 1 (FR1), and can only meet the requirements of basic V2X services due to the limited  bandwidth~\cite{3GPP_5Greq2020}. 6G V2X is expected to support a wide range of services, i.e., enhanced-V2X (e-V2X), with increasingly stringent requirements: $>\,$10 Gbps per link, $<\,$1 ms latency, and $<\,$1$\cdot$10$^{-9}$ reliability~\cite{6g}. To address these new demands, the 3GPP has launched 5G New Radio (NR) V2X Rel. 16~\cite{TR22886} and further enhancements from Rel. 17~\cite{rel17} are under discussion. The most significant innovation in Rel. 17 is the possibility of exploiting mmW frequencies (24.25 GHz-52.6 GHz), i.e., Frequency Range 2 (FR2), to support bandwidth-demanding applications. Furthermore, sub-THz (90-300 GHz) are being studied for 6G V2X communications~\cite{Gozalvez8887840}. 

Propagation at such high frequencies is subject to severe attenuation and mostly requires an LoS condition. Indeed, blockage is the most relevant limitation for the range covered and for communication reliability~\cite{TVT_mmw}. Massive Multiple-Input Multiple-Output (MIMO) systems are promising solutions for high-frequency communication, enabling a beam-type communication capable of achieving high antenna gain. Beam alignment and tracking are key issues, mostly in a dynamic environment as in vehicular scenarios, where frequent beam misalignments and blockages occur, with a detrimental effect on the system's performance~\cite{Rappaport:J13}. 

Beam blockage can be mitigated by relaying mechanisms~\cite{v2v_multihop_mmwave,TVT_self_org}, for example, from nearby CAVs, Road Side Units (RSU), or Intelligent Reflecting Surfaces (IRS)~\cite{TVT_irs}. Relay selection strategies can be grouped into two macro-categories: \textit{i)} centralized~\cite{gu2019auction, 7516562}: a network entity, such as Next Generation Base Station (gNB) or an elected vehicle, acts as a \textit{central body} (CB) that collects network state and defines the optimal relay selection; \textit{ii)} distributed~\cite{TVT_self_org,9072416}: each vehicle computes the ego optimal relay and attempts to directly interact with it. 
Relay selection algorithms require fresh and updated information regarding the network to optimally allocate the relaying resources. 
These operations introduce delays and require signaling. In the mmW vehicular scenario, the network information is outdated rapidly due to the high mobility \cite{9440703}, yielding the relay selection inefficient. Moreover, multi-hop links are subject to a higher blockage probability compared to direct links.
In this context, we design a novel criterion for dynamic LoS-map prediction and propose a novel framework for relay of opportunity selection to enable high-quality and stable V2X links. Relay selection is based on cooperative sensing to cope with LoS blockage conditions.

\subsection*{Related Works}

\begin{figure*}[t]
\centering
\subfloat[ \label{fig:Scen}]{\includegraphics[width=0.32\textwidth]{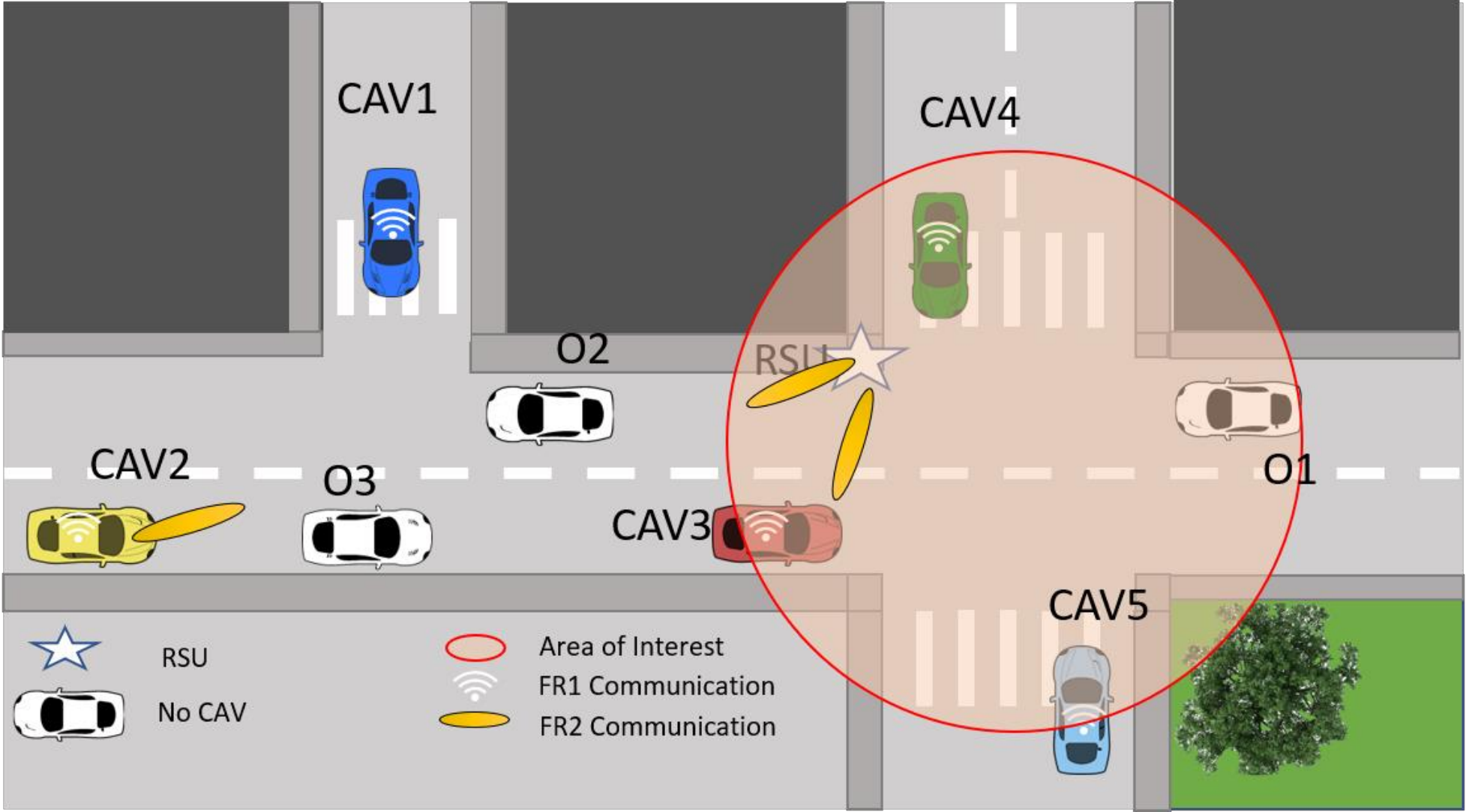}} \hspace{0.05 cm}
\subfloat[ \label{fig:Scen_sensing}]{\includegraphics[width=0.32\textwidth]{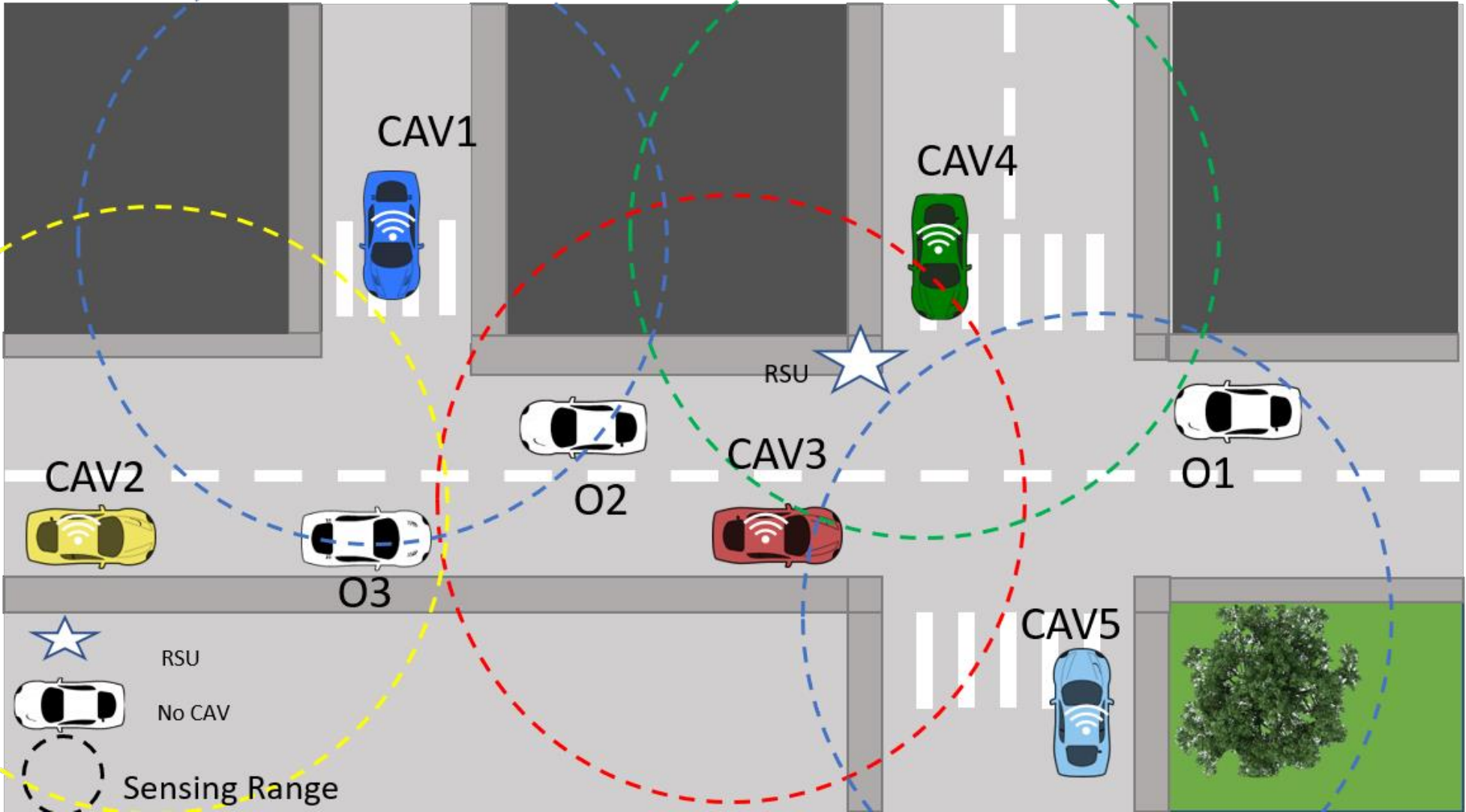}} \hspace{0.05 cm}
\subfloat[ \label{fig:Scen_pred}]{\includegraphics[width=0.32\textwidth]{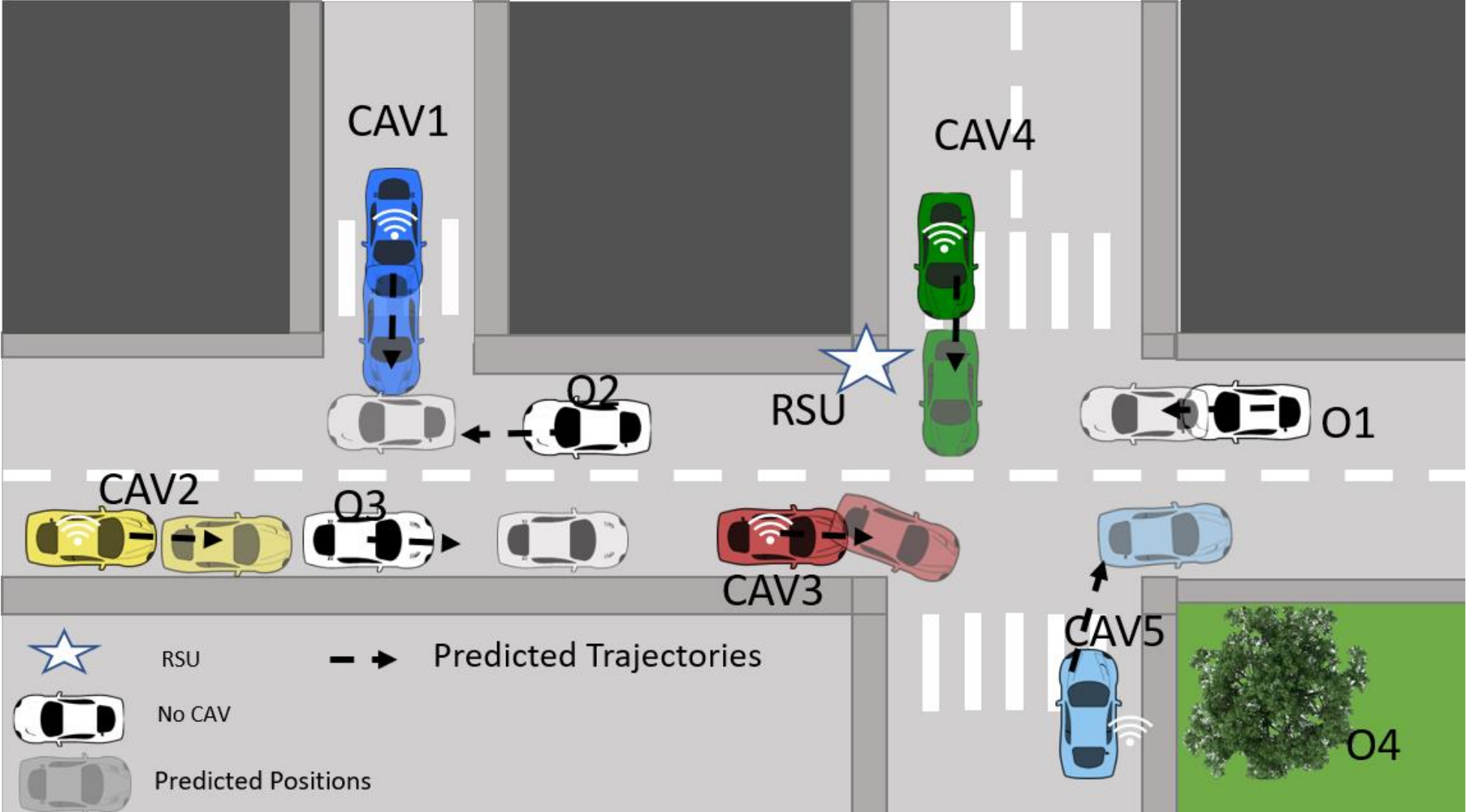}}
\caption{ V2X extended sensing between RV (e.g. yellow and blue car) and VoI (e.g. red and green cars).
In (a) the scenario at the collection phase instant, in (b) the Cooperative Awareness Signalling to enable cooperative sensing, and in (c) the predicted trajectories based on CA signalling and maneuvers- and physical-based models.}
\label{fig:scenario}
\end{figure*}

Many works in literature approach relay selection by abstracting the physical layer. The authors in~\cite{8662549} propose a distributed vehicular relay selection strategy assuming a Nakagami fading channel. A moving-zone-based architecture is proposed in~\cite{7516562}, where the CB is elected from the set of vehicles and it establishes a connectivity with the closest relay to the destination by using the Dijkstra algorithm.
In~\cite{guo2021predictor} the channel state information at the communication antenna is predicted based on a predictor antenna, which is a second antenna mounted on the CAV's front bumper. The presence of a predictor antenna can boost the relaying communication performance. However, the prediction time horizon depends on the speed of the CAV and the distance between predictor and communication antennas, which is limited by the size of the CAV.
Relaying in vehicular networks perspective has been extensively studied~\cite{soa2, performancecomp, soa4}. In particular, the authors in~\cite{performancecomp} proposed two relay schemes to extend the transmission range in a vehicular platooning application. The investigated approaches select the relay based on the geographic location information, showing a significant reduction in the outage probability compared to the single relay case. A multi-relay scheme is proposed in~\cite{ModelingAnalysisMulti-Relay}, where the base stations cooperate as relays to support vehicular communication, leveraging on the geographical location information. The authors in~\cite{PerformanceEvaluationRelaying} suggest the deployment of relay nodes in the network to support vehicular communication. Different relaying schemes are compared under different nodes density/deployment scenarios, considering instantaneous and average Signal to Noise Ratio (SNR) and data rate as selection bases.

In~\cite{MBR}, the authors assume that each vehicle
equipped with a GPS receiver can download the digital maps from the cloud. Their combination can be used to choose from the neighbors a relay node
that is strategically located in the perfect-reception zone.

In order to overcome the LoS blockage problem at mmW, in~\cite{map2} the authors propose
using neighbouring vehicles as relays to forward the blocked traffic flows. Specifically, a traffic-aware relay vehicle selection is
investigated, by combining the results of an analytic hierarchy process and coalitional game. A heuristic relay selection scheme is devised to select the relay vehicle with the best rationality degree. However, these works investigate solutions in simplified road traffic scenarios at sub-6GHz frequencies, without capturing the full complexity of dense vehicular environments and the harder propagation at mmW frequencies. Moreover, they assume the absence of old branded non connected cars that naturally interrupt the links.
Moreover, a common limitation of the works concerning the relay-assisted mmW V2X communications is the widespread use of stochastic models of the channel. This is very limiting in modelling the complex real-world scenario. For example, stochastic channels do not account for the impact on the system's performance of \textit{dynamic} blockers, such as moving vehicles.

In~\cite{mezzavilla2018end}, the authors use a Network Simulator 3 (NS-3)~\cite{ns3} mmW module and 5G Lena~\cite{patriciello2019e2e} for simulating V2I communication, focusing only on cellular-like and infrastructure-based deployments. 
Coverage, mobility, and blockage are studied in~\cite{wang2020demystifying} by exploiting a 3D ray-tracer to accurately reproduce the mmW channel propagation~\cite{lubke2020channel}. The simulation architecture achieves high performance in replicating the channel propagation model but its high complexity fails to capture the insight of the dynamics for relay scheduling design.

\subsection*{Contributions}

This paper is contextualized in the future 6G-enabled CAV applications/services where CAVs share sensor data to improve the perception of the surrounding environment. CAVs are equipped with two communication interfaces: mmW beam-type communication for high data rate and a sub-6 GHz for signalling and control~\cite{gozalvez} as depicted in Fig.~\ref{fig:Scen}. This paper's main contributions can be summarized as follows:
\begin{itemize} [wide]
    \item The proposed approach leverages on cooperative sensing to predict the LoS-map of the V2X network and to design a countermeasure to beam blockage. The static objects (e.g., buildings and trees) in the LoS-map can be retrieved either from the 3D maps downloaded from edge server or by the CAVs' on-board sensors. The trajectories of dynamic objects, such as non-cooperative vehicles, are derived by reshaping the motion prediction algorithm in~\cite{xie2017vehicle}. The vehicles' mobility and the uncertainty on the objects' position lead to dynamic LoS conditions, which depends on the traffic density, and impacts on the corresponding V2X connectivity.
    \item The distribution of the LoS-map is derived analytically for V2X scenarios. Instantaneous network connectivity for a dynamic vehicular mesh is evaluated in the conditions of no relaying (lower bound), and optimal relaying (upper bound), i.e., all CAVs can act as relays with unlimited resources. This enables the comparison of different relaying techniques, both centralized and distributed, under various link and traffic conditions.
    \item A unified simulation methodology of mmW V2X networks is proposed. The modelling of vehicle traffic over real road networks is obtained by using a combination of OpenStreetMap (OSM)~\cite{osm} and Simulation of Urban MObility (SUMO)~\cite{SUMO2018} software. Furthermore, the output of SUMO is processed by the Geometry-based, Efficient Propagation Model for Vehicle-to-Vehicle (GEMV$^{2}$)~\cite{GEMV2} software, in which the radio propagation has been adapted according to the latest 3GPP guidelines in~\cite{14rel}. All numerical results are based on this integrated simulation environment.
\end{itemize} 

\subsection*{Organization}
The rest of this paper is organized as follows. Section~II introduces the methodology, while Section~III deals with the network connectivity analysis. Section~IV gives an overview of centralized relay selection schemes, while Section~V focuses on the distributed relay selection schemes.
Section~VI describes the implemented simulation set-up with scenario definition and the CAVs' data sharing studied cases. Numerical results and complexity analysis are investigated in Section~VII. Lastly, Section~VIII summarizes and concludes the work.

\subsection*{Notation}
Bold upper and lower-case letters describe matrices and column vectors, respectively. $\left[\mathbf{{A}}\right]_{(i,j)}$ denotes the $(i,j)$ entry of matrix $\mathbf{A}$. Transposition and conjugate transposition are denoted with the operator $(\cdot)^{\mathrm{T}}$ and $(\cdot)^{\mathrm{H}}$, respectively. The determinant of matrix $\mathbf{A}$ is det$(\mathbf{A})$. $|\cdot|$ is the cardinality of a matrix or a set, i.e. its number of elements.
With $\mathbf{a}\sim\mathcal{CN}(\boldsymbol{\mu},\mathbf{C})$ we denote a multi-variate complex Gaussian random variable $\mathbf{a}$ with mean $\boldsymbol{\mu}$ and covariance $\mathbf{C}$. $\mathbf{I}_M$ stands for the identity matrix of dimensions $M\times M$, and $\mathbf{1}_M$ is the all-ones vector of dimensions $1 \times M$. 
The functions $f_X(x)$ and $F_X(x)$ define the probability density function (pdf) and cumulative density function (cdf) of a random variable $X$, respectively.
The notation ${m}^{(i,j)}$ is used to indicate a metric $m$ of the link between the $(i,j)$th communication couple, while ${p}^{i}$  associates the parameter $p$ to the $i$th CAV, RSU, and/or object. 

\section{Methodology}

This section introduces the proposed methodology for the selection of the relay of opportunity based on a prediction of the LoS-map. First, the upcoming network state is estimated by leveraging the static information and cooperative perception of CAVs. Then, the predicted links' availability, i.e., the probability that the SNR is higher than a predefined threshold, is used for relay selection schemes.

\begin{figure*}[t!] 
\centering
  \includegraphics[width=\textwidth]{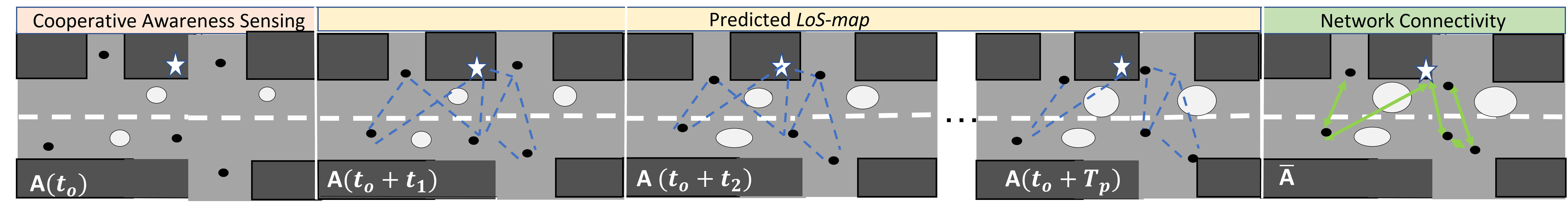}
  \caption{V2X network evolution while moving: after sensing, each CAV knows the others' state (black dots); the metrics based on predicted trajectories (black for CAV and white dots for old branded cars) are used to build an adjacency matrix $\mathbf{A}(t)$ for each time instant $t$. The entries of $\mathbf{A}(t)$ represent the service probability of the communication links involving CAVs and RSU, and it depends on the SNR.}
  \label{fig:graph}
\end{figure*}

\subsection{Study Case}

CAVs are assumed to be equipped with two communication interfaces: \textit{i)} a low data-rate and omnidirectional communication interface for Cooperative Awareness (CA) messages, dissemination and signalling (e.g., 5G NR in FR1); \textit{ii)} a high data-rate and beam-based communication interface for raw sensor data sharing (e.g., 5G NR in FR2 or 6G Sub-THz). Moreover, each CAV is assumed to share its planned trajectory with nearby CAVs.

Figure~\ref{fig:scenario} depicts the considered scenario with sketched beams. We consider V2X extended sensing as the 3GPP use case~\cite{14rel}. The CAVs exchange raw or nearly raw sensor data to achieve an augmented perception. More specifically, a Requesting Vehicle (RV) can establish direct access to the sensor of a Vehicle of Interest (VoI) after creating an Over-the-Air data bus, or virtual data bus~\cite{ChoVaGonDanBhaHea:J16} for sensor' data exchange. A CAV that is approaching a certain critical area, e.g., road intersection or roundabout, requires an augmented perception of the objects/obstacles present in the scene and their state to perform a safe manoeuvre. By referring to Fig.~\ref{fig:Scen}, we denote the critical area as the Area of Interest (AoI), the collaborating CAVs as VoI (e.g., green and red cars in Fig.~\ref{fig:scenario}a), and the approaching CAVs as RVs (e.g., blue and yellow cars in to Fig.~\ref{fig:scenario}a). 
The VoIs inside the intersection (or roundabout) possess an advantageous point of view of that road segment's traffic flow. The incoming vehicles (i.e., RVs), on the other hand, are entering a critical road area, and need to gather as much perceptive information as possible.

\subsection{Cooperative LoS-map Sensing} \label{sec:sens}

The on-board sensing of each CAV perceives the surrounding static and dynamic blockers at a specific time instant $t_0$. The mutual positioning is then exchanged using the FR1 communication interface to obtain the cooperative sensing of the objects. 
The signalling to accomplish cooperative sensing has been standardized by the Intelligent Transportation System (ITS) society for CAVs, mainly for safety-related use cases. According to 3GPP~\cite{CAMstandard}, vehicles can exploit three types of Cooperative Awareness (CA) signaling: Cooperative Awareness Message (CAM), Decentralized Environmental Notification Message (DENM), and Cooperative Perception Messages (CPM). CAMs enable mutual awareness between CAVs and detected obstacles, DENMs are event-based messages containing information related to a road hazard or abnormal traffic conditions, and CPMs contain self-information, such as the available on-board sensors, sensors range, sensors field of view, position, speed, heading, and size of the detected objects. 

In particular, for each detected object $k$ at time instant $t_0$, the $i$th CAV estimates its position $(\tilde{x}_i^{k}(t_0), \tilde{y}_i^{k}(t_0))$ and speed $\tilde{v}_i^{k}(t_0)$ along $x$ and $y$ directions of the Cartesian plane. These parameters define the state of the $k$th object, which is represented as the column vector
\begin{align} \label{eq:initial_error}
        \mathbf{\tilde{s}}^{k}_i (t_0) &= \left[\tilde{x}_i^{k} (t_0), \tilde{y}_i^{k} (t_0), \tilde{v}_{i,x}^{k} (t_0),  \tilde{v}_{i,y}^{k} (t_0)\right]^T 
\end{align} 
that due to the measurement uncertainty can be approximated as $\mathbf{\tilde{s}}^{k}_i (t_0)$$\,\sim\,$$\mathcal{N}\left(\mathbf{s}^{k}_{true} (t_0), \boldsymbol{\tilde{\Sigma}}_i^k (t_0)\right)$ where $\mathbf{s}^{k}_{true} (t_0)$ is the true state of the $k$th object~\cite{s21010200}.

The list of $K$ objects $\mathcal{L}_i $= $ \left\{{\mathbf{\tilde{s}}^1_i(t_0),\ldots,\tilde{\mathbf{s}}}^K_i(t_0)\right\}$ sensed by the $i$th CAV and their spatial footprint (see Remark 1) are signalled by leveraging on CA messages through the FR1 interface.

An estimate of the state associated with the $k$th object is obtained by each CAV (for a distributed architecture) after reaching cooperative consensus or by the CB (in case of a centralized architecture). 
Assuming the presence of $N_c$ CAVs, the initial state of the \textit{k}th object is approximated as
\begin{align} \label{eq:state_after_cons}
    \mathbf{\tilde{s}}^{k} (t_0)\,\sim\,\mathcal{N}\left( \mathbf{s}^{k}_{true} (t_0),  \boldsymbol{\tilde{\Sigma}}^k (t_0)\right  ),
\end{align}
where the covariance matrix is 
\begin{align}\label{eq:cov}
    \begin{split}
        \boldsymbol{\tilde{\Sigma}}^k(t_0)= \left(\sum_{i=1}^{N_c} \boldsymbol{\tilde{\Sigma}}^k_{\text{i}}(t_0)^{\,\,-1}\right)^{-1}.
    \end{split}
\end{align}

In the proposed framework, the CA signalling at FR1 is exploited to enable the communication at mmW frequency (e.g., in Fig.~\ref{fig:Scen_sensing}). In fact, by combining the CA-based information and the blockers states, each CAV can evaluate its static LoS-map at $t_0$ and predict the dynamic LoS-map in the interval $[t_0, t_0+T_p)$, estimating the LoS condition between each tuple $(i,j)$ of CAVs in the network over the prediction time window $T_p$. The LoS condition classification from 3GPP standards depends on the obstacle type; therefore, the link between the two CAVs can be classified as NLoSb/NLoSf in the case of building or foliage blockage, NLoSv in case of vehicle blockage, and as LoS if no blockage is affecting the link.

\begin{remark}
\normalfont 
Each CAV that measures the state of a $k$th object needs to estimate its spatial footprint.
For a non-connected autonomous vehicle (nCAV), its localization of the point  $(\tilde{x}_i^{k}(t_0), \tilde{y}_i^{k}(t_0))$ has an intrinsic uncertainty within the vehicle's shape. The uncertainty compounds the normal distribution of an estimate point~\cite{s21010200} and the vehicle's shape. The result can be approximated to a normal bivariate for high uncertainty. Thus, the footprint area of the vehicle can be modelled by an isotropic disc of radius $\sigma_i^k(t_0)$, whose value is taken into account in the initialization of the covariance matrix $\boldsymbol{\tilde{\Sigma}}_i^k (t_0)$\,=\,$\sigma_i^k(t_0)^2 \mathbf{I}$. These considerations allow us to adopt Gaussian models.
\end{remark}

\subsection{Prediction of the LoS-map} \label{sec:pred}

The cooperative sensing information collected makes it possible to predict the state of CAVs in the time window $T_p$. Many prediction strategies are discussed in literature~\cite{xie2017vehicle, soatti1, track, 8690640}. The proposed method revises and reshapes the work in~\cite{xie2017vehicle}, where the predicted state at time instant $\bar{t}$ is
\begin{align} \label{eq:pred_pos}
     \mathbf{\tilde{s}}^{k} (\bar{t}) = \mathbf{T}(\bar{t}) \mathbf{\tilde{s}}^{k} ({t}_0) + \mathbf{w}^{k} (\bar{t}), 
\end{align}
where
\begin{align}
    \mathbf{T}(\bar{t}) = \begin{bmatrix}
\mathbf{I}_2 & \bar{t}\,\mathbf{I}_2 \\
\mathbf{0}_{2 \times 2} & \mathbf{I}_2
\end{bmatrix}
\end{align}
is the transition matrix of the constant velocity model and $\mathbf{w}^k (\bar{t}) \sim \mathcal{N} \left(\mathbf{0}, \mathbf{Q}^k \left(\bar{t}\right)\right)$ is the prediction error, characterized by the covariance $\mathbf{Q^k(\bar{t})}$, which is derived by fitting the numerical results of~\cite{xie2017vehicle}. Note that the authors in~\cite{xie2017vehicle} consider the constant turn rate and acceleration model, which would require the position, speed, heading, turn rate, and acceleration, hence, it is more complex and it requires more parameters compared to the used model.

The proposed framework, depicted in Fig.~\ref{fig:graph}, exploits the predicted states to evaluate the dynamic evolution of LoS conditions among CAVs over the time window $T_p$. 

\subsection{Network Metrics Evaluation} \label{sec:metrics}

\begin{table*}[t] 
\centering
\caption{Direct component characterization of the path-loss models from 3GPP~\cite{14rel, itu, goldhirsh}}
\label{PL}
\begin{small}
\begin{tabular}{ | c | c | c | c |}
	\hline
	\textbf{Condition}  & \textbf{Mean value} ($A_{PL}(d)$) [dB] & \textbf{Symbol} & \textbf{Description} \\ \hline \hline
	LoS (highway) & $32.4+20\log_{10}(d)+20\log_{10}(f)$ & $A_{LoS}(d)$  & clear link visibility \\ 
	LoS (urban) & $38.77+16.7\log_{10}(d)+18.2\log_{10}(f)$ & \,  & \, \\ 
	NLoSb &  $36.85+30\log_{10}(d)+ 18.9\log_{10}(f)$  & $A_{NLoSb}(d)$ & buildings blockage \\ 
	NLoSf & $A_{LoS}(d) + A_{f}$  & $A_{NLoSf}(d)$ & foliage obstructions\\
	NLoSv & $A_{LoS}(d) + A_{v}$  & $A_{NLoSv}(d)$ & vehicular blockage \\ \hline
	\end{tabular}
\end{small}
\end{table*}

The QoS of the link $i-j$ can be characterized by its SNR $\gamma^{(i,j)}(t)$. 
For simplicity of notation, the indexes $(i,j)$ that identify a certain link will be omitted since the discussion can be generalized for any tuple of CAVs. 
The time evolution of the SNR $\gamma$ depends on configurable parameters (such as transmission power and antenna gains) and non-configurable ones (such as the distance between transmitting and receiving CAVs and blocking obstacles). Static obstacles are inferred from the digital map and CA signalling, while the state of dynamic nCAVs is predicted in \eqref{eq:pred_pos}. The predicted motions make the LoS-map dynamically evolving over time (see Fig.~\ref{fig:graph}). 

We assume that CAVs are equipped with $N_a = N_v \times N_c$ cylindrical antenna arrays on their rooftop, where $N_v$ is the number of uniform circular arrays, characterized by $N_c$ uniformly spaced antenna elements each as defined in~\cite{MIZMIZI2021100402} (any other array configuration can be used and it needs a straightforward adaptation).
The SNR at time $\bar{t}$ in decibel scale is
\begin{align} \label{eq:snr}
    \gamma(\bar{t})&=P_{Tx}\,+\,2G_b\,-PL(\bar{t})\,-\,P_n = \nonumber \\
    &= \gamma_0(\bar{t})\,-\,PL(\bar{t}) \,\,\ \left[\text{dB}\right]
\end{align}
where $P_{Tx}$ is the transmitted power, $G_b$ is the antenna gain, $P_n$ is the noise power, and path-loss $PL(\bar{t}) \sim \mathcal{N}(A_{LOS}(d(\bar{t})),\sigma^2_{sh})$ contains log-normal shadowing with variance $\sigma^2_{sh}$ and $A_{LoS}(d(\bar{t}))$, where $d$ refers to the distance, in Table~\ref{PL}. The proper PL model is selected based on the 3GPP recommendations in \cite{3GPP_5Greq2020, 14rel} given the current propagation environment, i.e., urban, rural, or highway. CAVs have access to their position and map information that are combined to obtain context information about the current propagation environment. The randomness of the SNR is analyzed next.

\begin{remark}
\normalfont 
In the case of multiple blockages, the order of relevance in the $PL(\bar{t})$ computation depends on the blocker with the highest attenuation~\cite{14rel,3gppNRSidelink}, in order (from Table~\ref{PL}): NLoSb, NLoSf, and lastly NLoSv. For example, if the LoS is blocked by a building and a vehicle, the used path-loss model is NLoSb.
\end{remark}

\subsection{Link SNR Statistics}
\begin{figure} [t!]
    \centering
    \includegraphics[width=0.75\columnwidth]{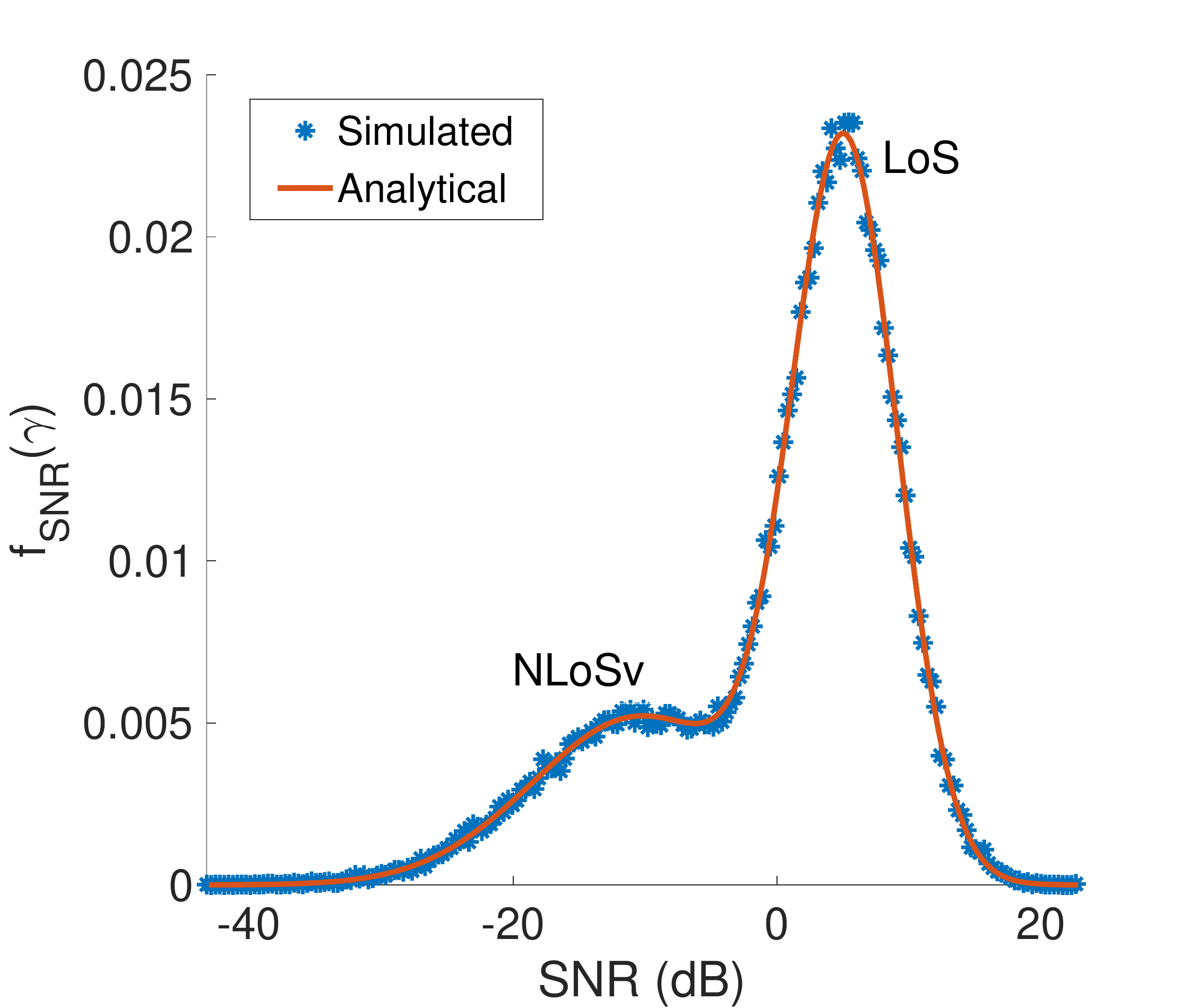}
    \caption{Analytical and numerical SNR $\gamma$ pdf}
    \label{fig:sim_teo}
\end{figure}

\begin{figure}[!t] 
\centering
\subfloat[Scenario of uncertainty between LoS and NLoSv \label{fig:unc}]{\includegraphics[width=0.75\columnwidth]{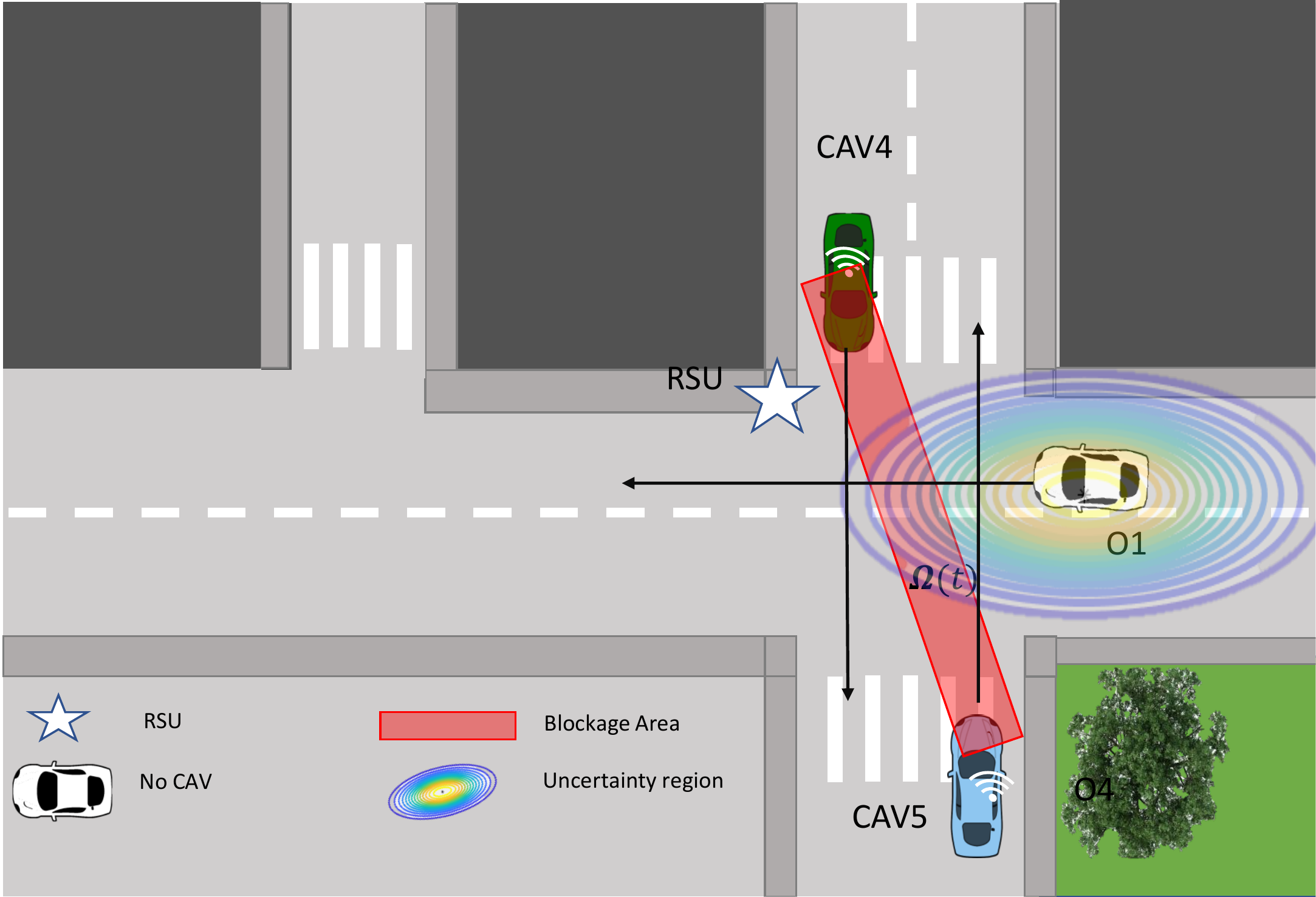}}

\subfloat[SNR pdf versus prediction time\label{fig:ac}]{ \includegraphics[width=0.75\columnwidth ]{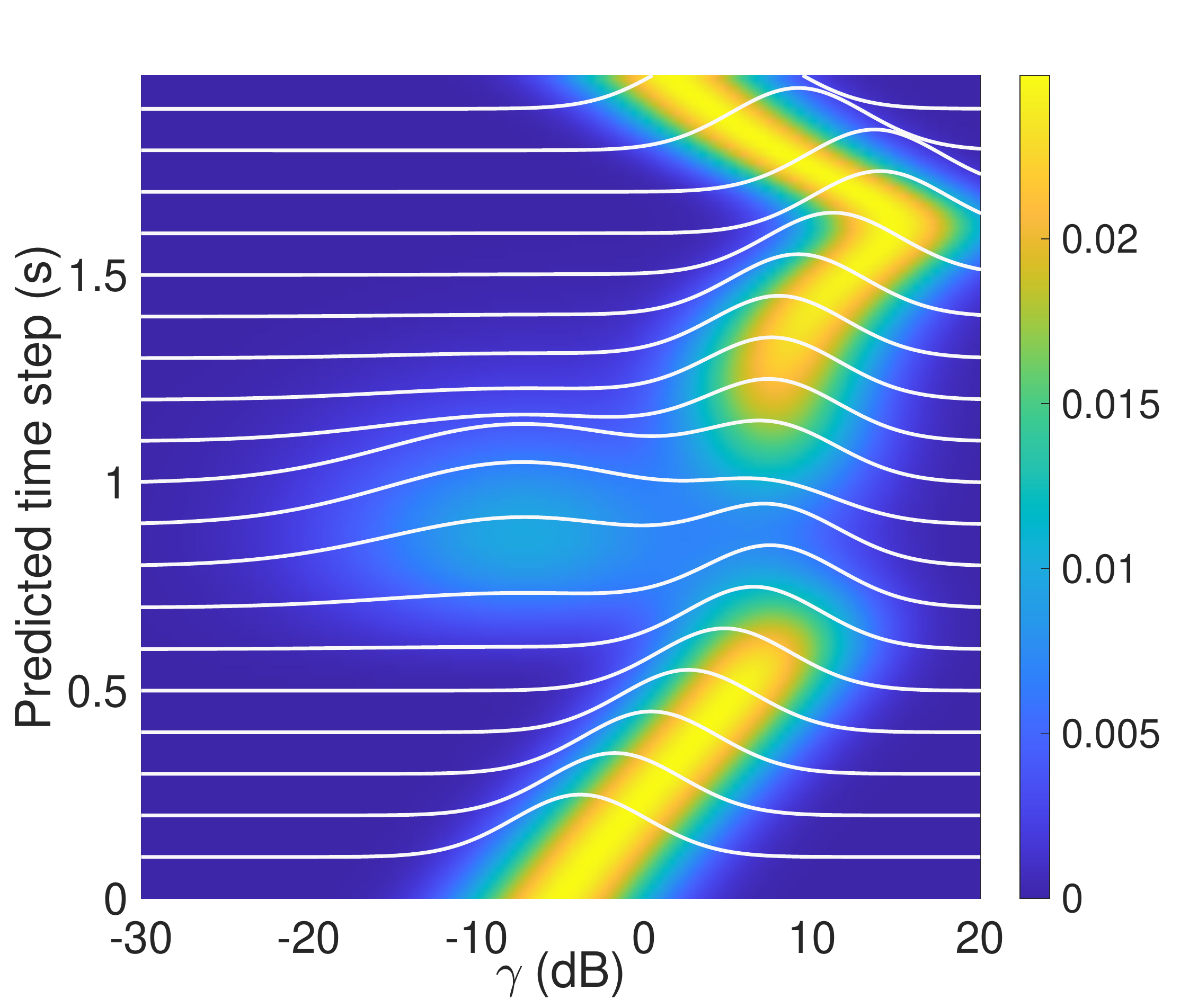}}

\caption{CAVs link pair SNR pdf evolution according to the predicted positions in case of uncertainty between LoS and NLoSv.}
\label{fig:acc}
\end{figure}
In our framework, current and future positions of all CAVs are considered known with negligible error. This assumption is based on the localization and tracking systems of self-driving CAVs that combine multiple sensors to reach sub-metre position accuracy, including the individual footprints with exact shapes~\cite{rahman2018improving, TVT_path_planning}. Furthermore, self-driving cars know their future trajectory as it is typically self planned~\cite{negar}. 
The SNR $\gamma$ distribution, based on the LoS condition, is mainly affected by the uncertainty of the nCAVs positions and shadowing. 

In the remaining section, we evaluate the pdf of the SNR in closed form. Two illustrative cases are considered below for the dynamic LoS-map:
\begin{itemize} [wide]
    \item \textbf{absence of nCAVs}: the SNR distribution (in decibel scale) $f_{SNR}(\gamma)$ for the LoS and NLoSb/f cases is
    \begin{align}\label{eq:snr_no_nCAVs}
      f_{SNR} (\gamma) =  \mathcal{N} ( \gamma_0 - A_{PL}(d) , \sigma^2_{sh}),
    \end{align}
    with the variance $\sigma^2_{sh}$ related to the shadowing.
    In case of LoS and NLoSb the term $A_{PL}(d)$ is reported in Tab.~\ref{PL}, while for NLoSf, the path-loss $PL$ in \eqref{eq:snr} contains an additive random term
    $X_{extra} \sim \mathcal{N}(A_{f}, \sigma_{f}^2)$.
    The extra attenuation due to the foliage $A_f$ is computed using the concept of mean excess loss derived in~\cite{goldhirsh} and the overall path-loss distribution results in $PL \sim \mathcal{N}(A_{LoS}(d)+A_{f}, \sigma_{f}^2+\sigma_{sh}^2)$.
    \item \textbf{presence of nCAVs}: the SNR distribution $f_{SNR}(\gamma)$ depends on the shadowing component and the probability of blockage due to nCAVs. The extra attenuation for vehicular blockage $A_v$ depends on the number of blocking vehicles
    ~\cite{14rel}. The path-loss distribution is evaluated similarly to NLoSf case. The SNR pdf $f_{SNR}(\gamma)$ can be computed as
    \begin{align} \label{eq:snr_si_nCAVs}\footnotesize
        f_{\text{SNR}}(\gamma) = P_{LoS} f_{LoS}(\gamma) + \sum_{k_b = 1}^{ N_{b}} P_{NLoSv}^{(k_b)} f_{NLoSv}^{(k_b)}(\gamma),
    \end{align}
    where the term $P_{LoS}$ is the probability of LoS condition, $f_{LoS}(\gamma)$ is the related SNR pdf, $P_{NLoSv}^{(k_b)}$ is the probability of having $k_b$ nCAVs blockers out of $N_b$ nCAVs, and $f_{NLoSv}^{(k_b)}(\gamma)$ is the related SNR pdf. The pdf in \eqref{eq:snr_si_nCAVs} is a mixture of Gaussians. $P_{LoS}$ and  $P_{NLoSv}^{(k_b)}$ are complementary: $P_{LoS}+\sum_{k_b = 1}^{ N_{b}} P_{NLoSv}^{(k_b)} = 1$. Their analytical derivation, together with the LoS and NLoSv SNR pdfs, are given in Appendix A. Figure~\ref{fig:sim_teo} shows the analytical and simulated behaviour of the SNR pdf in \eqref{eq:snr_si_nCAVs} for single blockage, i.e., $k_b=N_b=1$.
\end{itemize}

Figure~\ref{fig:unc} illustrates a realistic use case where the link between two CAVs can be obstructed by an nCAV (O1). Its SNR distribution varies according to the vehicles' relative motion directions (black arrows). The pdf of the SNR for every prediction time instant $\bar{t}$ is reported in Fig.~\ref{fig:ac}b. It can be seen that, as soon as the nCAV uncertainty region is expected to intersect the blockage area (red polygon), the SNR drops as shown by the slices at $t\in(0.6,1.2)$ s, and after the nCAV departure the SNR rises again.

\section{Predicted Network Connectivity} \label{sec:connectivity}

Once the SNR distribution of each link associated with the LoS-map has been predicted, the adjacency matrix of CAVs' network is used to compute the connectivity. The entries of the adjacency matrix (say links $i-j$) are defined by the probability that SNR $\gamma^{(i,j)}({t})$ is higher than a threshold $\gamma_{\text{th}}$, which is given by the minimum QoS according to the specific V2X service.  
The adjacency matrix is then used as input of the relay selection strategies.

\subsection{Network Adjacency Matrix} \label{sec:graph}

The predicted pdfs of SNR $\gamma^{(i,j)}(t)$ are used to derive the adjacency matrix $\mathbf{A}(t) \in \mathbb{R}^{N_c \times N_c}$ of the CAVs' network in Fig~\ref{fig:graph}.
The entries of the matrix $\mathbf{A}(t)$ are the service probabilities that evolve with the LoS-map (or equivalently, with the motion of the vehicles):  
\begin{align}\label{eq:servProb}
 \left[\mathbf{{A}}\right]_{(i,j)}(t)&=\text{Pr} {\{ \gamma^{(i,j)}(t)}>\gamma_{\text{th}} \}=  
  F_{\mathrm{SNR}^{i,j}}(\gamma_{\text{th}}),
\end{align}
with $i,j$\,=\,$1,\ldots,N_c$ and $i \neq j$, and the cdf of the SNR $F_{\mathrm{SNR}}(\gamma_{\text{th}})$ follows from \eqref{eq:snr_si_nCAVs}, as derived in Appendix A.
The network performance for link-selection, as described in Sec.~IV, is based on the service probabilities in \eqref{eq:servProb} over the communication interval $(t_0, t_0+T_p]$, which is in turn discretized with sampling interval $T_s$. 
Averaging the service probabilities  \eqref{eq:servProb} over $N_s$\,=\,$T_p/T_s$ samples of the prediction window, we obtain the average \textit{link availability} that is
\begin{align} \label{eq:link_ava}
      \left[\mathbf{\bar{A}}\right]_{(i,j)}= \frac{1}{N_s}\sum_{n=0}^{N_s-1} { \left[\mathbf{{A}}\right]_{(i,j)}(t_0+nT_s)}.
\end{align}
The matrix of the link availabilities $\mathbf{\bar{A}}$ is symmetrical, and it is the key metric for the relay selection problem (see Secs.~IV~and~V).

\subsection{Connectivity Analysis}

\begin{figure} [t!]
    \centering

\subfloat[\label{fig: highway}]{ \includegraphics[width=0.4\columnwidth]{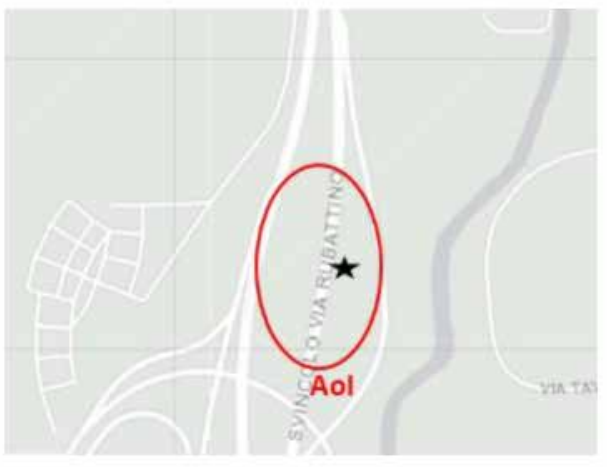}}
    \subfloat[ \label{fig:1st_con_high}]{\includegraphics[width=0.6\columnwidth]{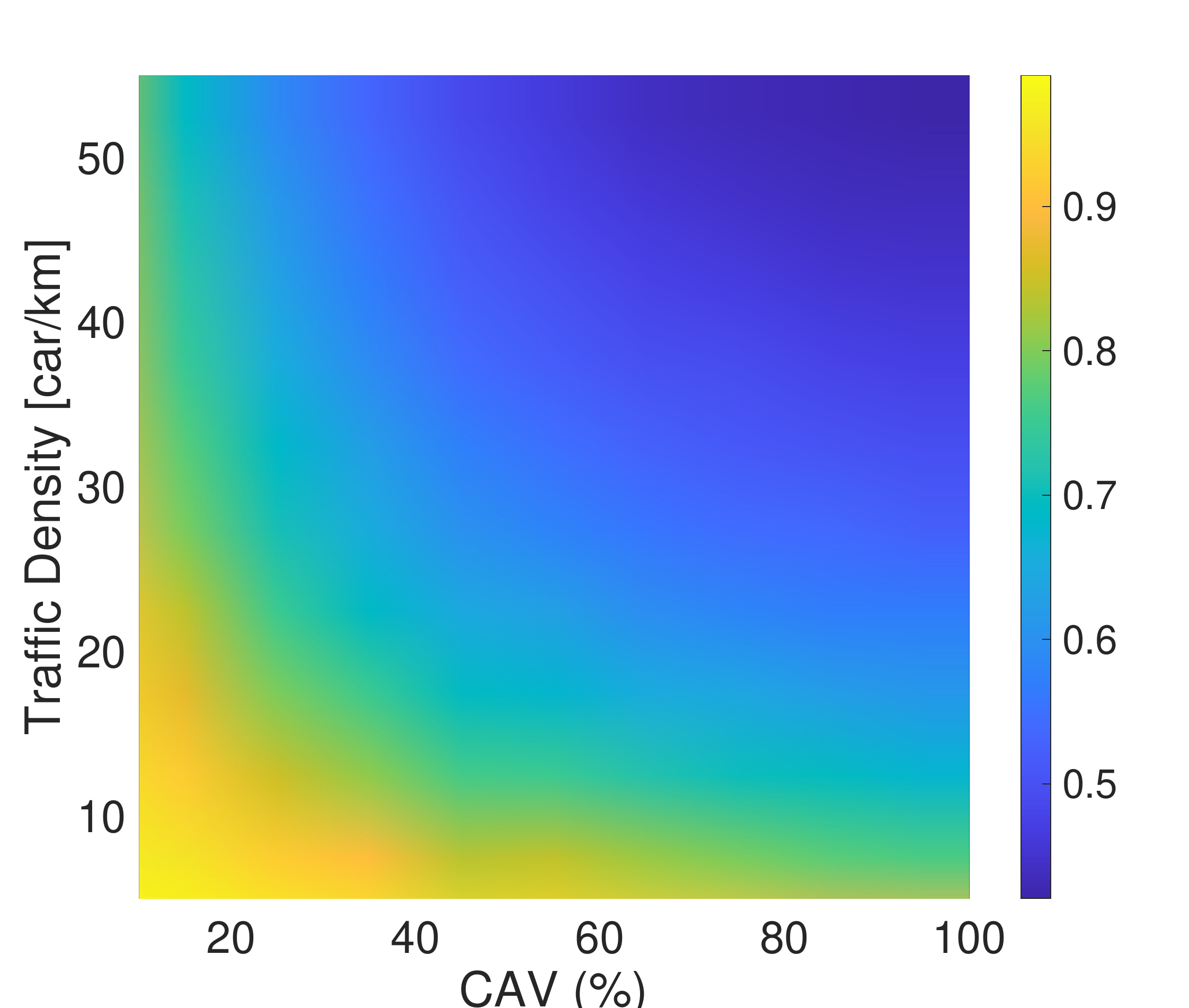}}
    
    \caption{(a) Simulated highway scenario and (b) numerical first order network connectivity  versus CAV percentage $\rho_c$ in percentage for different traffic density condition.}
    \label{fig:graph_con}
\end{figure}

The network connectivity is a metric to evaluate any need for relay nodes. Let the metric $\alpha(\mathbf{\bar{A}})$ be the measure of the connectivity of the network, $\alpha(\mathbf{\bar{A}})$ is related to the density of nCAVs and determines a-priori the reliability of the vehicles configuration and mesh of CAVs connectivity. 
The connectivity can be evaluated by calculating the Laplacian matrix~\cite{jamakovic2007relationship}
\begin{align}\label{eq:laplacian}
    \mathbf{\Delta}\stackrel{def}{=} \mathbf{D}-\mathbf{\bar{A}},
\end{align}
where the elements of $\mathbf{D}$ are the vertex outdegrees 
\begin{align}
    \left[\mathbf{D}\right]_{(i,i)} = \sum_{j \neq i}    \left[\mathbf{\bar{A}}\right]_{(i,j)}.
\end{align}
The second smallest eigenvalue $\lambda_2$ of  $\mathbf{\Delta}$ defines the network connectivity $\alpha(\bar{\mathbf{A}})$\,=\,$\lambda_2$, which is $\alpha(\bar{\mathbf{A}}) > 0$ in case of connected network. 
The second-order connectivity determines the network's robustness by considering two-hop links connecting two vehicles without any constraints on the relaying capabilities.
It is obtained by substituting $\mathbf{\bar{A}}$ with $ \mathbf{\bar{A}}_2$ in \eqref{eq:laplacian}, that is defined as
\begin{align}\label{eq:2ordAvailMatrix}
    \mathbf{\bar{A}}_2 =  \mathbf{\bar{A}}^2 + \mathbf{\bar{A}}.
\end{align}
This can be interpreted as a performance upper-bound for the relay schemes' connectivity capability. 

We study the first order connectivity in the highway scenario in Fig.~\ref{fig: highway} 
since it is characterized by simpler topology and mobility trajectories with respect to the urban environment. 
The simulation is performed by varying the traffic density and the percentage of CAV with $\gamma_{th}$\,=\,$10$dB. 
The V2X network based only on direct links lacks of robustness when the traffic density (veh/km) grows and for high percentage of CAVs (Fig.~\ref{fig:1st_con_high}). This is due to the impact of dynamic blockage, 
outlining the need for proactive  relay selection strategies. 
These considerations are also eligible for intersection and roundabout scenarios, where relaying schemes are more needful because of the harsher propagation conditions.

\subsection{Relay Selection}

The relay selection task can be formulated as an assignment problem, which belongs to a class of optimization problems. The goal is to maximize the network connectivity by assigning relays to RVs/VoIs based on the average availability under the constraint of limited relaying resources. In this regard, the relay selection task can be expressed as
\begin{subequations}
    \begin{alignat}{2} \label{eq:centralized_opt_problem}
        \max_{\mathbf{x}} & \quad \sum_l \sum_r \left[\mathbf{\hat{A}}_2\right]_{(l,r)} \left[\mathbf{X}\right]_{l,r}\\
         \text{such that} & \quad \sum_l \left[\mathbf{X}\right]_{l,r} \leq R^{r}_c\,\, \forall r\\
        & \quad \sum_r \left[\mathbf{X}\right]_{l,r} \leq 1\,\, \forall l,
    \end{alignat}
\end{subequations}
where $l$ refers to the link between the $i$th RV and the $j$th VoI, $r$ is the relay index, and $\left[\mathbf{\hat{A}}_2\right]_{(l,r)}$ = $\left[\mathbf{\bar{A}}\right]_{(i,r)}\left[\mathbf{\bar{A}}\right]_{(j,r)}$ is the average availability of the relayed link. The assignment variable $\left[\mathbf{X}\right]_{l,r}$ is 
\begin{equation}
\left[\mathbf{X}\right]_{l,r} = \begin{cases} 1, \;\; \text{if $r$ is assigned to $l$}, \\
0, \;\;  \text{otherwise} .\\
\end{cases}
\end{equation}

The problem in \eqref{eq:centralized_opt_problem} can be solved through centralized or distributed approach, as proposed below.

\section{Predictive Centralized Approach}\label{sec:central}

Centralized link selection strategies require overall knowledge of the network state to address the following issues: \textit{which CAVs request a link? Which CAVs are available to act as relay nodes? How CAVs are connected in the network?} 
Typically, a CB, such as an elected CAV, the base station, or the RSU, through periodical FR1 communications, is responsible for \textit{(i)} collecting the network information, link requests, and relaying capabilities, \textit{(ii)} applying the selection algorithm and \textit{(iii)} communicating the decision to the involved communication parties. 
In this section, we first introduce the communications signalling required for information gathering and distribution of decisions. Later, we discuss the proposed algorithms for link selection. 

\begin{figure} [b!]
    \centering
    \includegraphics[width=0.95\columnwidth]{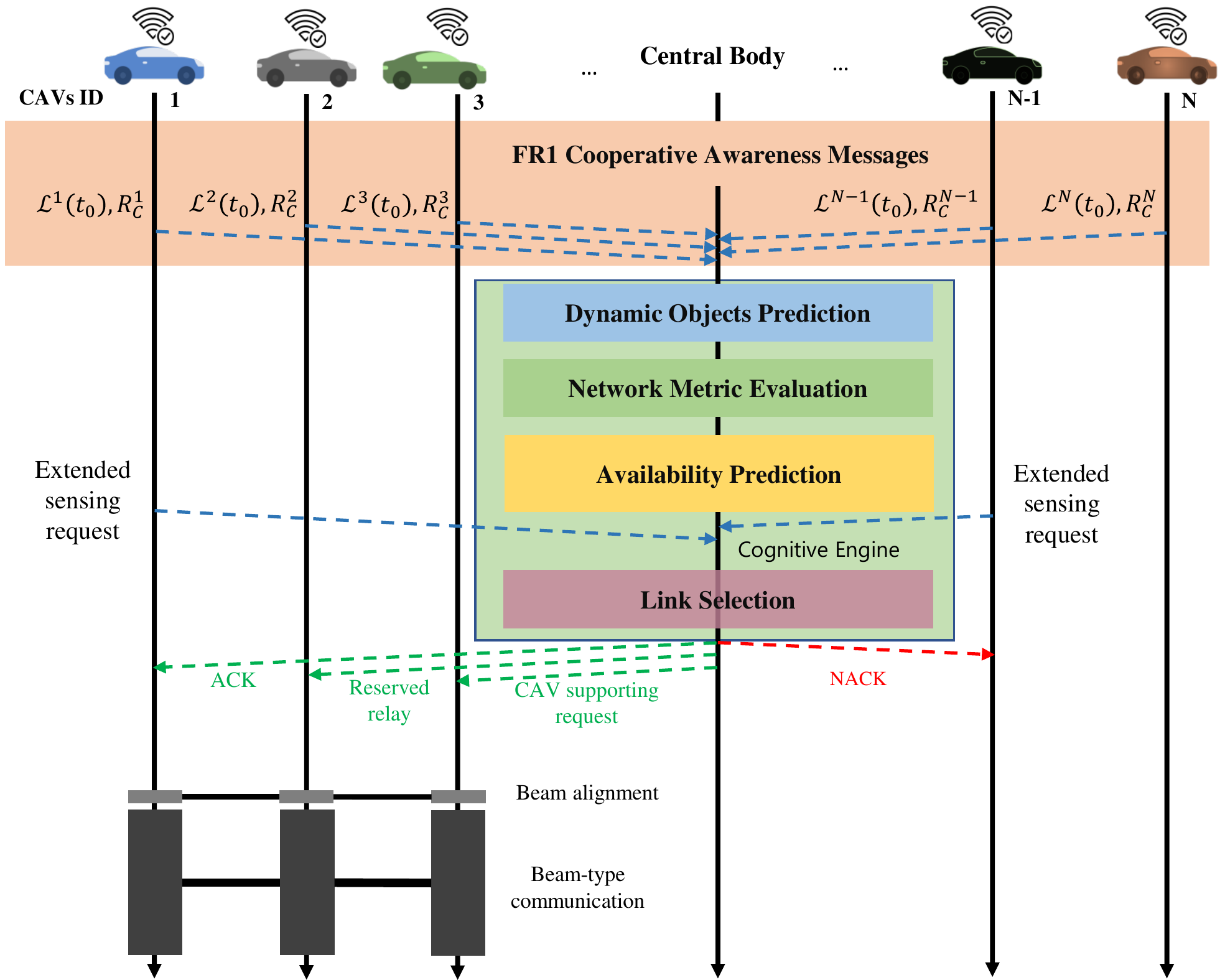}
    \caption{Centralized approach time flow example. CAVs ID1 and IDN-1 are RVs, ID3 is a VoI, and ID2 is a relay. Dashed lines are FR1 communication links, while continuous lines represent beam-type communication, e.g. FR2.}
    \label{fig:central}
\end{figure}

\subsection{Centralized Protocol}

Figure~\ref{fig:central} depicts an example time flow for the centralized protocol, i.e., the messages and procedures supporting link selection. In particular: 
\begin{enumerate}[wide]
    \item \textbf{Cooperative Sensing:} the $i$th CAV, after the sensing operation, obtains a list of sensed objects $\mathcal{L}_i$ that is sent to the CB together with the CAV relaying capability $R_C^i$, which represents the maximum number of links it can support as a relay. 
    \item \textbf{Dynamic Objects Prediction:} the CB, after combining the received lists $\mathcal{L}_i$ for each $k$th object (e.g., pedestrian, CAV, and nCAV), predicts the dynamic evolution of the sensed nearby objects over a predefined time window $T_p$, as discussed Sec.~\ref{sec:pred}.
    
    \item \textbf{Network Metric Evaluation:}  CB evaluates, for each predicted time step $\bar{t}$, the $\gamma^{(i,j)}(\bar{t})$ between each tuple $(i,j)$ of CAVs in the network. The $\gamma^{(i,j)}(\bar{t})$ is computed using \eqref{eq:snr}.
    
    \item \textbf{Prediction of Dynamic Adjacency Matrix} is derived by the CB based on the previously estimated SNRs. The entries of the adjacency matrix are computed based on \eqref{eq:servProb}.
    
    \item \textbf{Link Selection:} CAVs (i.e., RVs) that plan to cross an AoI could send a request for extended sensing to the CB, as depicted in Fig.~\ref{fig:central}. The CB collects the received requests and provides the optimal link based on the predicted link availability in \eqref{eq:link_ava} and on the link selection algorithm. If this link exists, the CB sends an acknowledgment (ACK) to the RV (otherwise, it sends a NACK), a request to the CAV supporting extended sensing, and a relay reservation request. Finally, the CB sends the necessary information to establish an FR2 communication link to the involved parties, as in Fig.~\ref{fig:central}.
\end{enumerate}

The centralized protocol is executed periodically so that the information obtained through cooperative sensing is updated, and, consequently, the network metrics and the predicted link availability.

\subsection{Centralized Link Selection}

The optimization problem in \eqref{eq:centralized_opt_problem} is linear and, therefore, convex. There are several methods in the literature to solve it, and they differ mainly in complexity. For the centralized scheme, we consider the following approaches:
\begin{itemize}
    \item \textbf{Exhaustive search (ES):} is the most intuitive method of finding an optimal assignment, which is achieved by searching all the possible arrangements. Considering $N$ RVs and $M$ relays, each with $R_c^r$ resources, the complexity of the ES method is $\mathcal{O}(\,(MR_c^r)^{N}\,)$.
    \item \textbf{Hungarian Game (HG):} is a combinatorial optimization algorithm that solves the assignment problem with polynomial time \cite{burkard1999linear}. To apply the HG algorithm, we need to reformulate the assignment problem in \eqref{eq:centralized_opt_problem} in matrix form:
    \begin{equation}
        \min_{\mathbf{L},\mathbf{R}} \mathbf{L}\;\mathbf{C}\;\mathbf{R}
    \end{equation}
	where $\mathbf{L}$, $\mathbf{R}$ are permutation matrices and $\mathbf{C}\in \mathbb{R}^{N\times MR_c^r}$ is a matrix whose elements are defined as
	\begin{equation}\label{eq:hung}
	    \mathbf{C} = -\mathbf{\hat{A}}_2.
	\end{equation}
	The complexity of the HG algorithm has been investigated in ~\cite{wang2007study}, and it is $\mathcal{O}\left(\,\left[\,\text{min}(N,MR_c^r)\left|\mathbf{C}\right|\,\right]\right)$. For a dense urban scenario we usually have $N \geq MR_c^r$. Therefore, the computational complexity can be expressed as $\mathcal{O}(\,N^2MR_c^r\,)$.
\end{itemize}

\section{Predictive Distributed Approach}\label{sec:Distributed}

Distributed link selection approaches do not rely on a CB. CAVs individually compute the optimal link locally and compete for the available relays. 
\begin{figure} [b!]
    \centering
    \includegraphics[width=0.95\columnwidth]{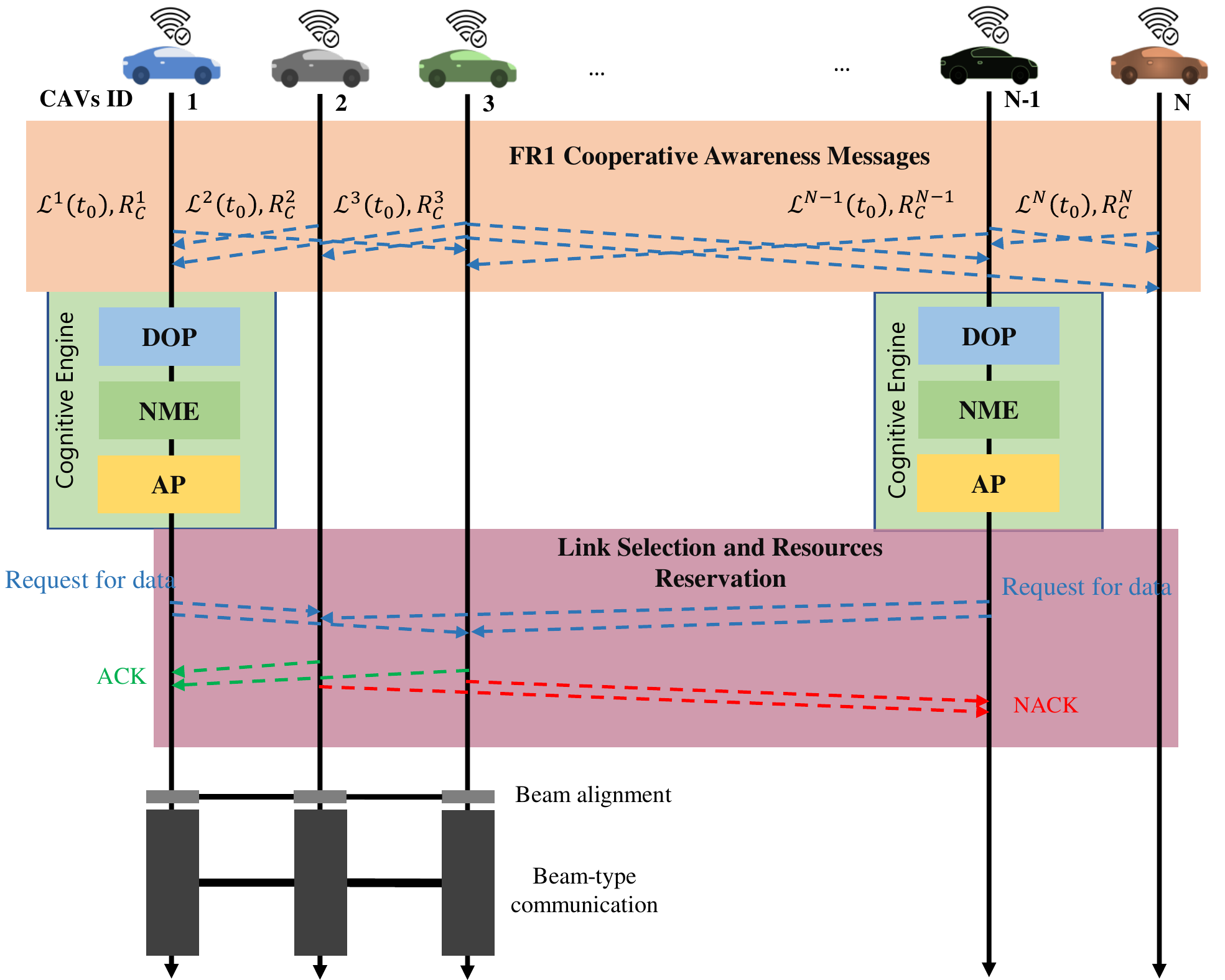}
    \caption{Distributed approach time flow example. CAVs ID1 and IDN-1 are RVs, ID3 is a VoI, and ID2 is a relay. Dashed lines are FR1 communication links, while continuous lines represents beam-type communication, e.g., FR2.}
    \label{fig:Distrib}
\end{figure}

\subsection{Distributed Protocol}

The time flow of the proposed distributed architecture (Fig.~\ref{fig:Distrib}) is similar to the centralized one. The main differences are related to cooperative sensing and link selection that are executed locally by the RVs only. In particular:
\begin{enumerate}[wide]
    \item \textbf{Cooperative sensing:} differently from the centralized method, each CAV shares with neighbouring CAVs the list of detected objects $\mathcal{L}_i$ and its relaying capability $R_C^i$. 
    \item \textbf{Link selection and resources reservation:} each RV executes the link selection algorithm based on the predicted link availabilities in \eqref{eq:link_ava}. The RVs send resources reservations requests to the selected relays, as depicted in Fig.~\ref{fig:Distrib}. The relays, based on their available relaying capabilities, reply to the RVs with positive or negative feedbacks.
\end{enumerate}
As for the centralized case, the distributed protocol is executed periodically to update the cooperative sensing information and the predicted link availability.

\subsection{Distributed Link Selection}
The optimization problem in \eqref{eq:centralized_opt_problem} regarding the maximization of the network connectivity is here solved in a distributed manner. Two distributed algorithms are detailed below:
\begin{itemize}[wide]
    \item \textbf{First Come First Served (FCFS):} is based on the \textit{first input first output} scheduling strategy. The RV sends the request for resource reservation to the $r$th relay associated to the maximum link availability $\left[\mathbf{\hat{A}}_2\right]_{(l,r)}$. If the relay's capability is sufficient, it sends a ACK, otherwise a NACK. The relays are assigned sequentially, therefore,  computational complexity is $\mathcal{O}(M)$.
    
    \item \textbf{Alternating Direction Method of Multipliers (ADMM):} solves a global optimization problem by splitting it into smaller local subproblems, which iteratively attempt to find the optimal global solution~\cite{boyd2011distributed}. Using ADMM involves creating copies of the optimization variables that are shared by different RVs. The optimization problem in \eqref{eq:centralized_opt_problem} can be reformulated as
    \begin{subequations}
    \begin{alignat}{2} \label{eq:admm_opt_problem}
        \min_{\mathbf{x}} & \quad \sum_l  \mathbf{x}_l\mathbf{c}_l^T \\
         \text{such that} & \quad \sum_l \mathbf{x}_l \leq R^{r}_c\mathbf{1}_M\\
        & \quad \mathbf{x}_l\mathbf{1}_M\leq 1\,\, \forall l,
    \end{alignat}
    \end{subequations}
    where the optimization variable is $\mathbf{x}_l$ $\in \mathbb{R}^{1\times M} $ is the $l$th row of the selection matrix $\mathbf{X}$ in \eqref{eq:centralized_opt_problem}, $\mathbf{c}_l \in \mathbb{R}^{1\times M}$ is the $l$th row of the matrix $\mathbf{C}$ in \eqref{eq:hung}. The ADMM algorithm solves the problem in \eqref{eq:admm_opt_problem} iteratively using the method of Lagrangian multiplier~\cite{boyd2011distributed}. 
    
    \textbf{Example:} To make an exemplary case and to clarify the ADMM application, let us assume $N=2$. The optimization problem in \eqref{eq:admm_opt_problem} simplifies to 
    \begin{subequations}
    \begin{alignat}{2} \label{eq:admm_example}
        \min_{\mathbf{x}_1, \mathbf{x}_2} & \quad f(\mathbf{x}_1)+f(\mathbf{x}_2) \\
         \text{such that} & \label{eq:cons1_admm_example}\quad  \mathbf{x}_1 + \mathbf{x}_2 \leq R^{r}_c\mathbf{1}_M \\
        &\label{eq:cons2_admm_example}\quad \mathbf{x}_1\mathbf{1}_M \leq 1\\ 
        &\label{eq:cons3_admm_example}\quad \mathbf{x}_2\mathbf{1}_M  \leq 1, 
    \end{alignat}
    \end{subequations}
    where the objective function $f(\mathbf{x}) = \mathbf{x}\mathbf{c}^\mathrm{T}$ makes the optimization convex. The augmented Lagrangian incorporates the constraints \eqref{eq:cons1_admm_example},\eqref{eq:cons2_admm_example},\eqref{eq:cons3_admm_example}:
    \begin{align} 
        \begin{split}
        \label{eq:lagran}
         L_{\rho}&(\mathbf{x}_1,\mathbf{x}_2, \boldsymbol{v}_1, \boldsymbol{v}_2, \boldsymbol{v}_3) = f(\mathbf{x}_1) + f(\mathbf{x}_2) \\ 
         &+ \boldsymbol{v}_1^T(\mathbf{x}_1 + \mathbf{x}_2-R^{r}_c\mathbf{1}_M) + \boldsymbol{v}_2^T(\mathbf{x}_1\mathbf{1}_M-1) \\
         &+ \boldsymbol{v}_3^T(\mathbf{x}_2\mathbf{1}_M-1) +\frac{\rho}{2}||\mathbf{x}_1 +  \mathbf{x}_2-R^{r}_c\mathbf{1}_M||^2_{2} \\
         &+\frac{\rho}{2}||\mathbf{x}_1\mathbf{1}_M -1 ||^2_{2}+\frac{\rho}{2}||\mathbf{x}_2\mathbf{1}_M -1 ||^2_{2},
        \end{split}
    \end{align}
    The optimization variables $\mathbf{x}_1$ and $\mathbf{x}_2$ are updated 
    \begin{subequations}
    \begin{alignat}{2}
        \mathbf{x}_1^{(k+1)} &:= \min_{\mathbf{x}_1}L_{\rho}(\mathbf{x}_1,\mathbf{x}_2^{(k)},\boldsymbol{v}_1^{(k)}, \boldsymbol{v}_2^{(k)}),\\
        \mathbf{x}_2^{(k+1)} &:= \min_{\mathbf{x}_2}L_{\rho}(\mathbf{x}_1^{(k+1)},\mathbf{x}_2,\boldsymbol{v}_1^{(k)}, \boldsymbol{v}_3^{(k)}),
    \end{alignat}
    \end{subequations}
    and the Lagrangian multipliers $\boldsymbol{v}_1$, $\boldsymbol{v}_2$, and $\boldsymbol{v}_3$ update are 
    \begin{subequations}
    \begin{alignat}{2}
        \boldsymbol{v}_1^{(k+1)} &:= \boldsymbol{v}_1^{(k)} + \rho(\mathbf{x}_1^{(k+1)} + \boldsymbol{x}_2^{(k+1)}-R^{r}_c\mathbf{1}_M), \\
        \boldsymbol{v}_2^{(k+1)} &:= \boldsymbol{v}_2^{(k)} + \rho(\mathbf{x}_1^{(k+1)}\mathbf{1}_M -1), \\
        \boldsymbol{v}_3^{(k+1)} &:= \boldsymbol{v}_3^{(k)} + \rho(\mathbf{x}_2^{(k+1)}\mathbf{1}_M -1).
    \end{alignat}
    \end{subequations}

    The updated variables $\mathbf{x}_1^{(k+1)}$, $\mathbf{x}_2^{(k+1)}$, $\boldsymbol{v}_1^{(k+1)}$, $\boldsymbol{v}_2^{(k+1)}$, and $\boldsymbol{v}_3^{(k+1)}$ are shared among the RVs through signaling in FR1. The stopping criterion depends both on the residual convergence to zero, and on the maximum number of iterations that can be fixed. Therefore, the ADMM complexity is $\mathcal{O}(T_k(NM+N+1))$, where $T_k$ is the number of iterations considered, which depends on the maximum tolerable delay induced by signaling.
\end{itemize}
\begin{table}[!t] 
\centering 
\caption{Summary of computational complexity}\label{tab:complexity}
\begin{tabular}{ | c | c | }
	\hline
	\textbf{Algorithm}  & \textbf{Complexity} \\ \hline\hline
	Exhaustive Search   & $\mathcal{O}(\,(MR_c^r)^N\,)$  \\ 
	Hungarian Game      & \hspace{-0.2cm} $ O(N^2MR_c^r)$ \\ 
	FCFS                & $\mathcal{O}(M)$  \\ 
	ADMM                & $\mathcal{O}(T_k(NM+N+1))$  \\ \hline\hline
\end{tabular}
\end{table}

\section{Simulations Framework} \label{sec:setup}

\begin{figure}[b!] 
  \includegraphics[width=0.85\columnwidth]{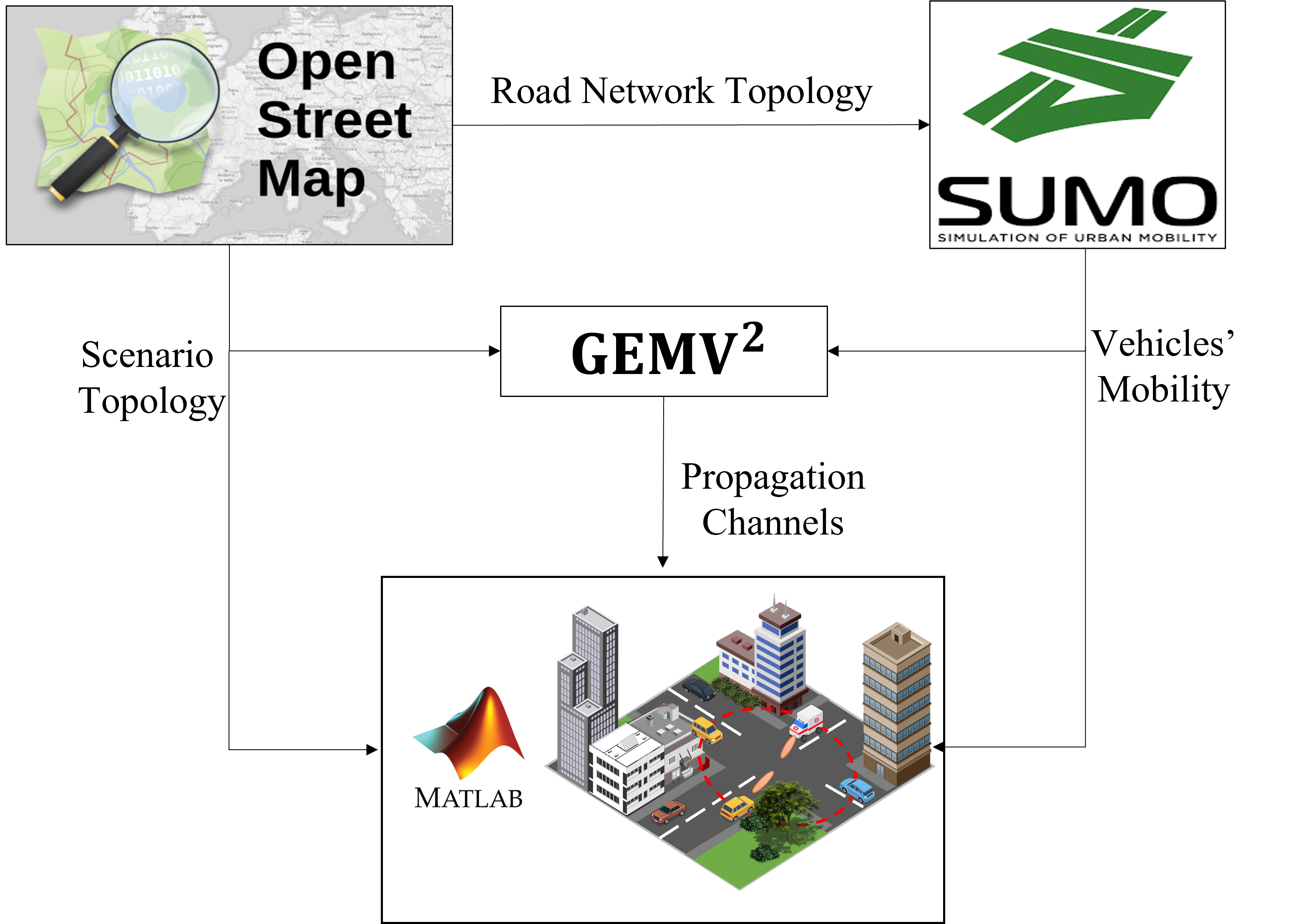}
  \caption{Simulation Methodology} \label{fig:meth}
\end{figure}

The mmW V2X communication is characterized by highly dynamic and sparse propagation channels~\cite{Rappaport6387266,Rappaport6834753}, subject to blockage by nCAVs, buildings, and trees. 
To better capture the system's complexity, the proposed framework combines several tools to simulate a realistic scenario, as depicted in Fig.~\ref{fig:meth}. In particular, the simulation environments, i.e., road network topology, buildings, and vegetation, are obtained from OSM~\cite{osm}. Realistic vehicular mobility is simulated by SUMO~\cite{SUMO2018}, and the GEMV$^2$~\cite{GEMV2} software generates the wireless channel in Matlab.

This section describes how the software tools are combined and set for numerical validation. Table~\ref{matlabparameters} reports the main  parameters used for the communications by following the 3GPP recommendations in~\cite{TS38101}.

\begin{table}[b!] 
\centering
\caption{Simulation parameters}\label{matlabparameters}
\footnotesize
\begin{tabular}{ | c | c | p{42mm} |}
	\hline 
	\textbf{Parameter}  & \textbf{Value} & \textbf{Description} \\ \hline\hline
	$P_{Tx}$    & $10$ dBm & Maximum transmitted Power\\ 
	$P_{{n}}$   & $-85.5$ dBm & Noise power \\
    $f^1_c$     & $5.2$ GHz & FR1 carrier frequency \\ 
	$f^2_c$     & $28$ GHz & FR2 carrier frequency \\ 
	$N_a$       &  $4\times16$ & Number of antenna elements\\
	$T_p$       & $1$ s & Predicted time window\\ 
	$R^{r}_c$	& $2$ & Maximum number of relay requests granted by a CAV and RSU \\  \hline \hline
	\end{tabular}
\end{table}

\subsection{Simulated Scenarios} \label{sec:sim_scen}
\begin{figure}[!t] 
\centering
\subfloat[Intersection \label{fig:intersection}]{\includegraphics[width=0.49\columnwidth]{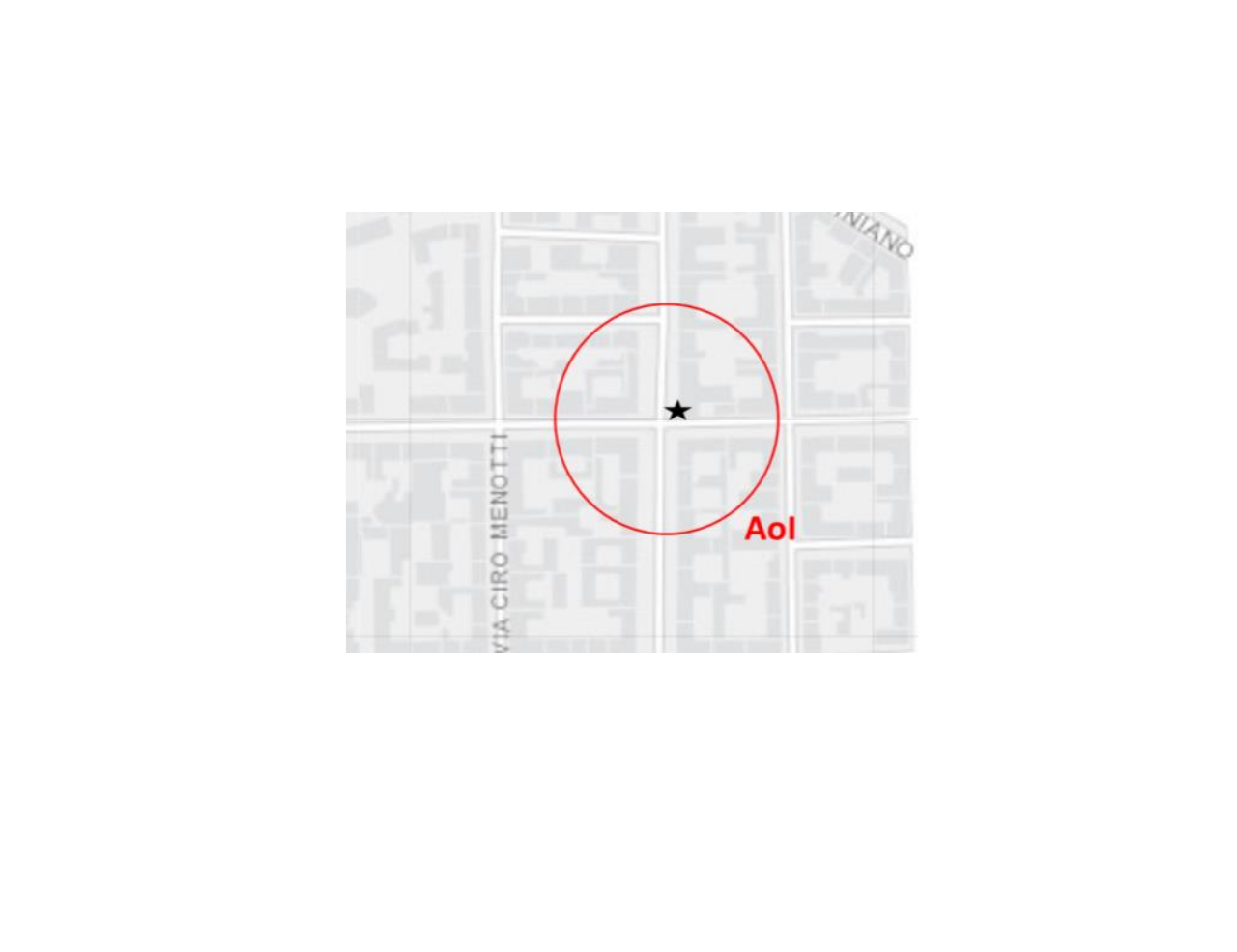}}
\subfloat[Roundabout \label{fig:roundabout}]{ \includegraphics[width=0.49\columnwidth ]{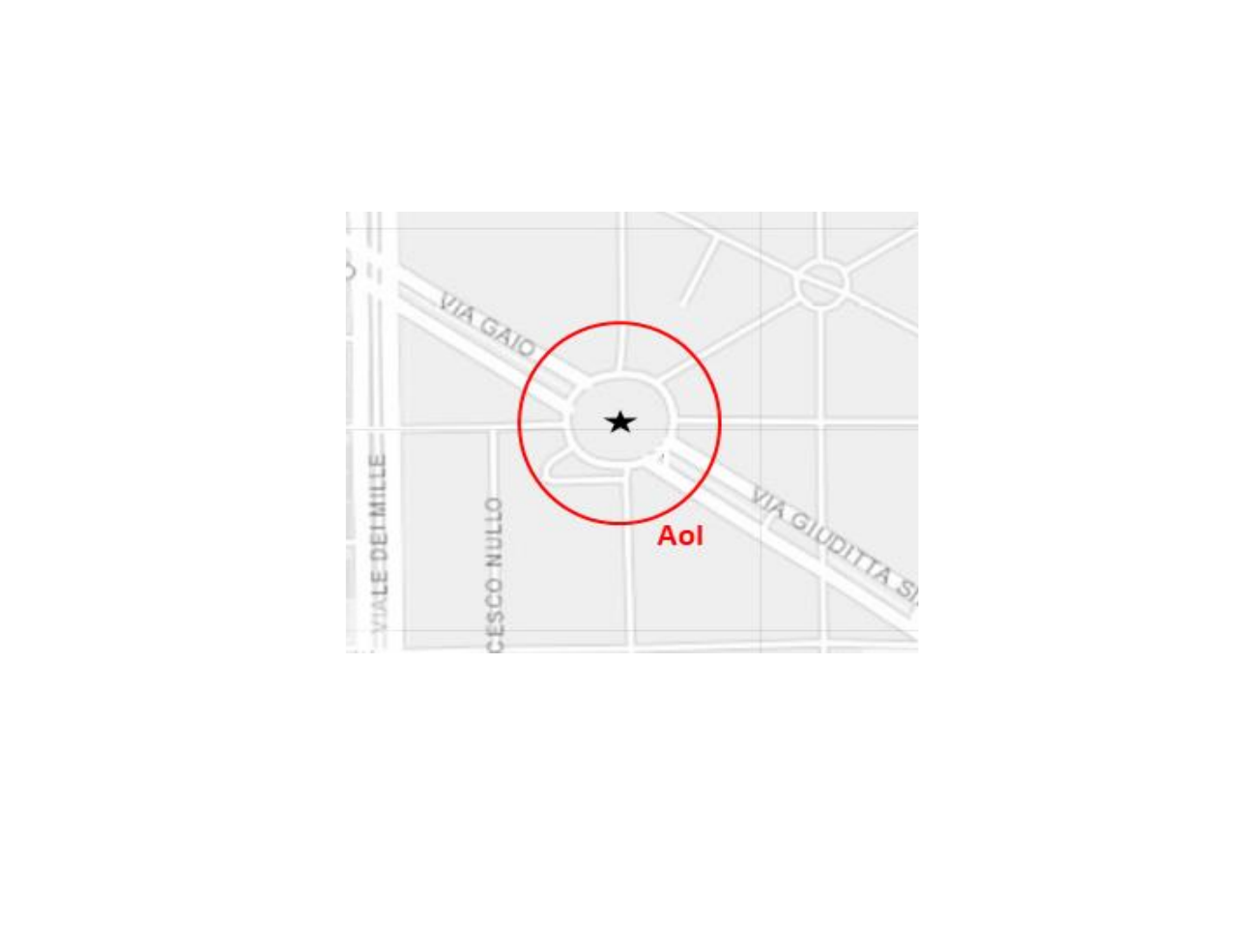}}
\caption{Simulated urban blind intersection (a) and roundabout (b).}
\label{fig:scenarios}
\end{figure}

A blind intersection (Fig.~\ref{fig:intersection}) and a roundabout scenario (Fig.~\ref{fig:roundabout}) are deemed to evaluate the performances of the proposed relay selection schemes. The selected areas are located around the city of Milan, Italy.
For each scenario, the AoI represents the critical area in the road environment. The CAVs, whose sensors' field of view is within the AoI are the VoIs. The range of the AoI (red circle) in the intersection and roundabout is 70 m and 90 m, respectively, and depends on the road patterns.
The RSUs are deployed in strategic positions, e.g., in the centre of the roundabout (see Fig.~\ref{fig:scenarios}), to assist the communication.

\subsection{Vehicles Topology and Mobility}

Simulation of vehicles' mobility is performed by SUMO software that allows for realistic traffic data generation. 
The \textit{road network}, used as input to SUMO, is generated from the selected OSM scenario.  
The output is processed in MATLAB together with the OSM scenario. Two vehicular density are simulated: medium and high with an average of 50 veh/km and 70 veh/km, respectively.

\subsection{V2X Communications Channel}

GEMV$^2$ \cite{GEMV2} is an open-source solution for V2X channel modelling with good accuracy and low complexity. It evaluates geometrically the propagation conditions and reflections using the outlines of buildings and foliage, provided by OSM, and the vehicles' ones, provided by SUMO. 
The propagation parameters used in the considered scenarios are defined according to the recommendations of 3GPP and International Telecommunications Union (ITU). In particular, we select the relative permittivity of ground, buildings, and vehicles based on the ITU measurements in~\cite{ITU-R2040}. The PL models for the propagation conditions are defined in Table~\ref{PL}.
GEMV$^{2}$ outputs the space-time features of the propagation channel, i.e., small and large scale channel components, delays, angles of departure, and angles of arrivals, which are used in \eqref{eq:snr} to evaluate the SNR for each link pair.

\section{Numerical Results}  \label{sec:results}

\begin{figure}[!b] 
\centering

\subfloat[ \label{fig:Con_versus_snr_Inter_medium}]{ \includegraphics[width=0.90\columnwidth]{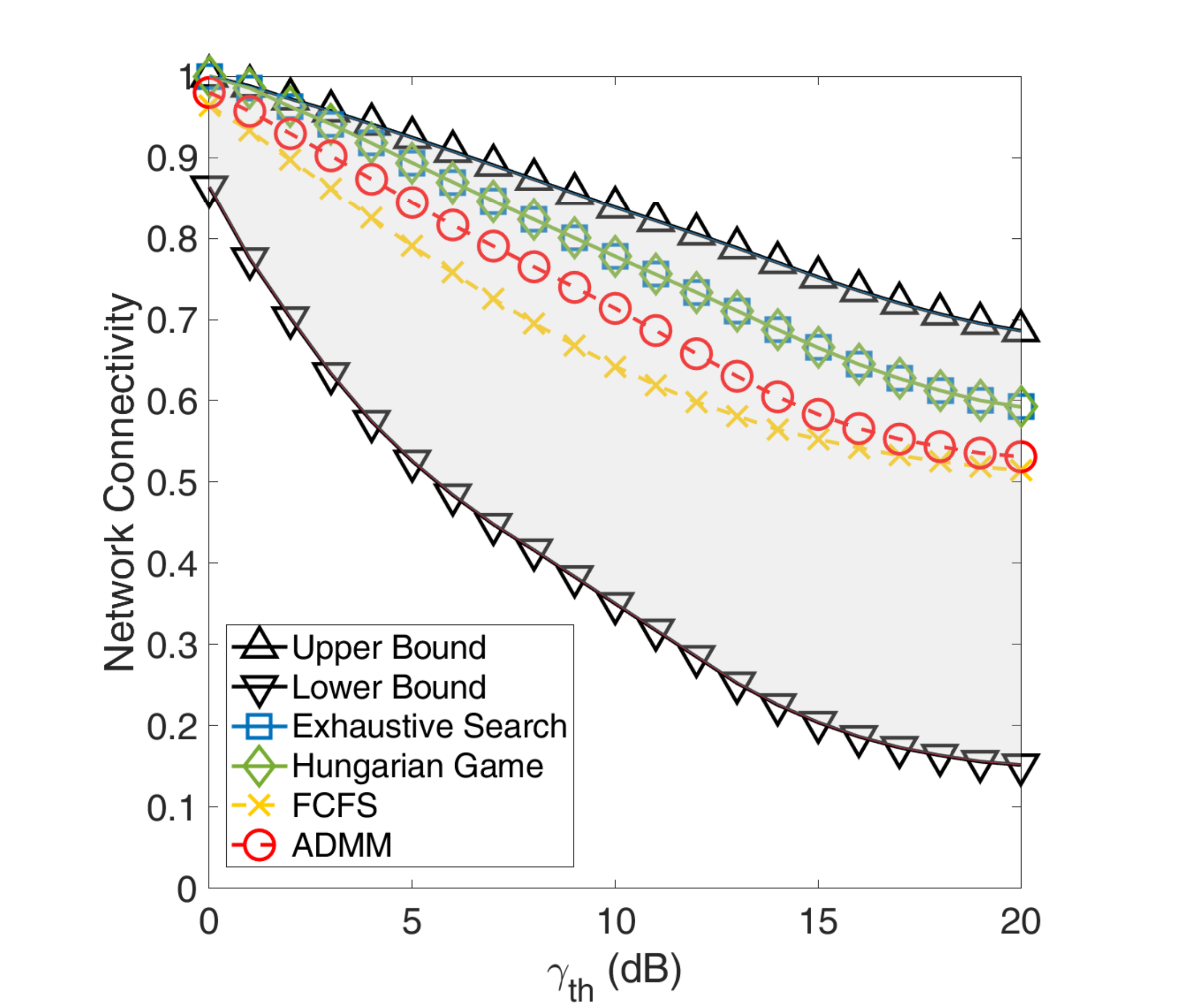}}\\

\subfloat[ \label{fig:Con_versus_snr_Inter_high}]{     \includegraphics[width=0.90\columnwidth]{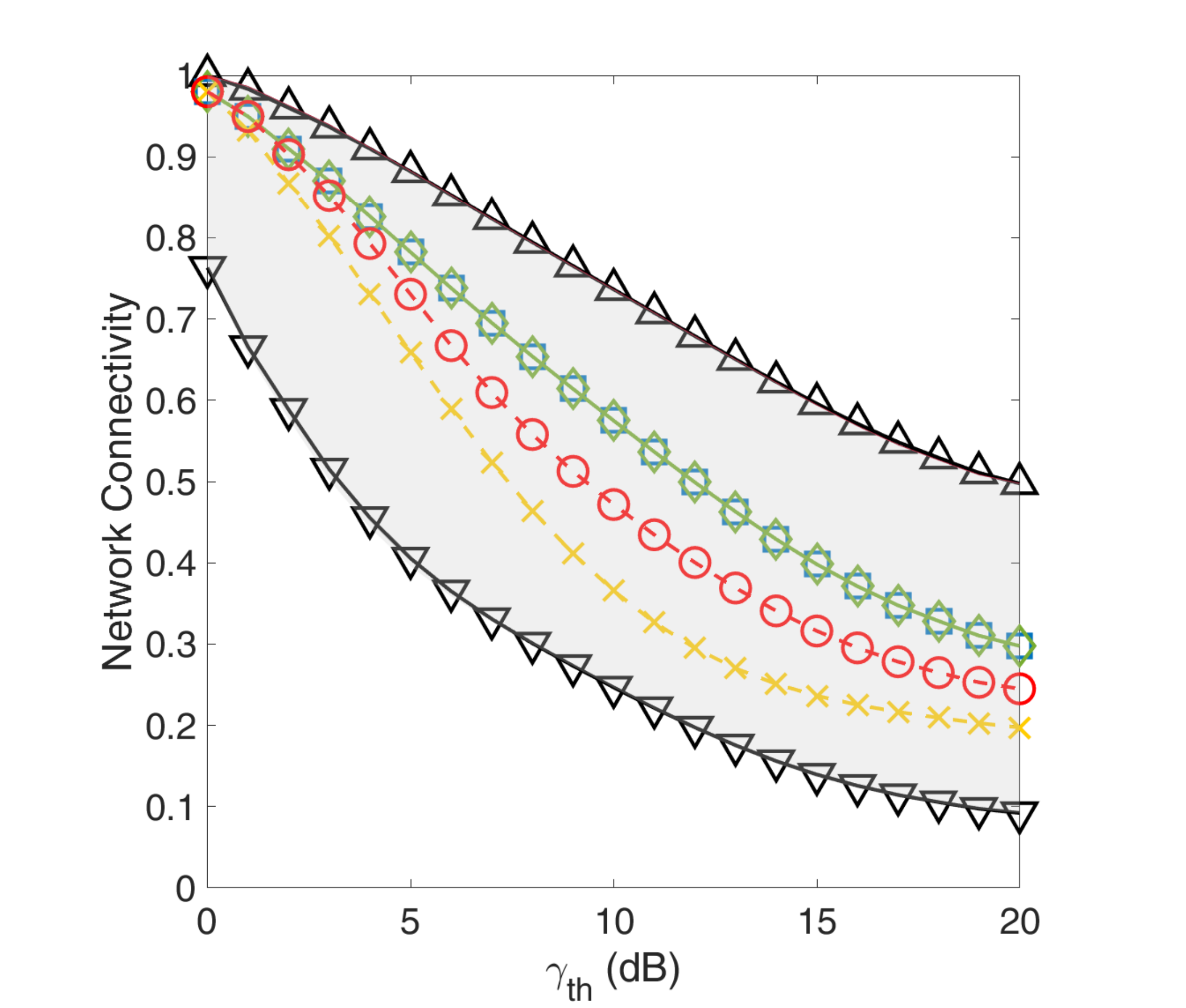}}

\caption{Network connectivity versus SNR $\gamma_{th}$ in urban intersection: (a) medium and (b) high vehicular density for all CAVs.}
\label{fig:connect_int}
\end{figure}

\begin{figure}[!b] 
\centering
\subfloat[ \label{fig:Con_versus_snr_round_medium}]{ \includegraphics[width=0.90\columnwidth]{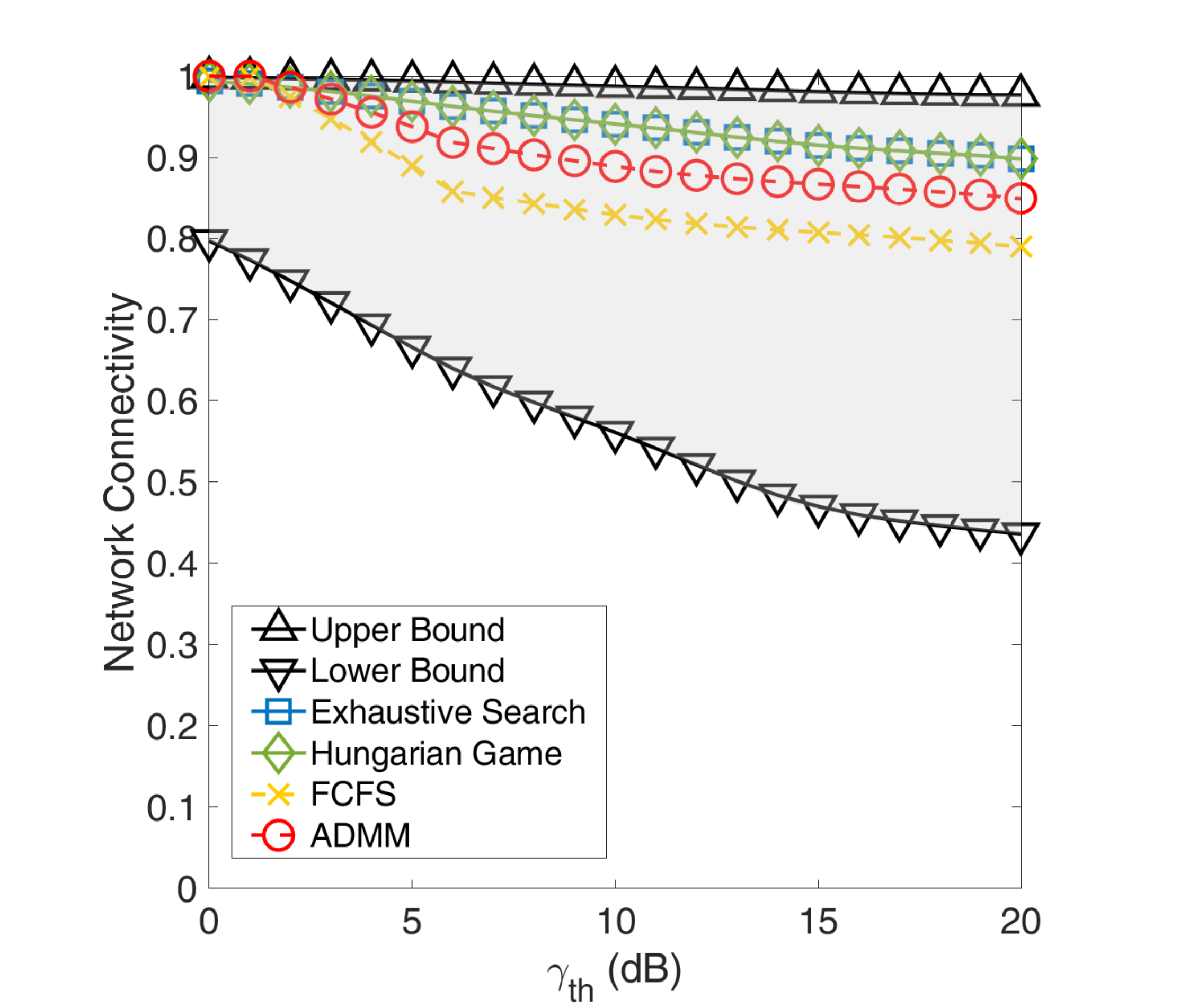}}\\
\subfloat[ \label{fig:Con_versus_snr_round_high}]{     \includegraphics[width=0.90\columnwidth]{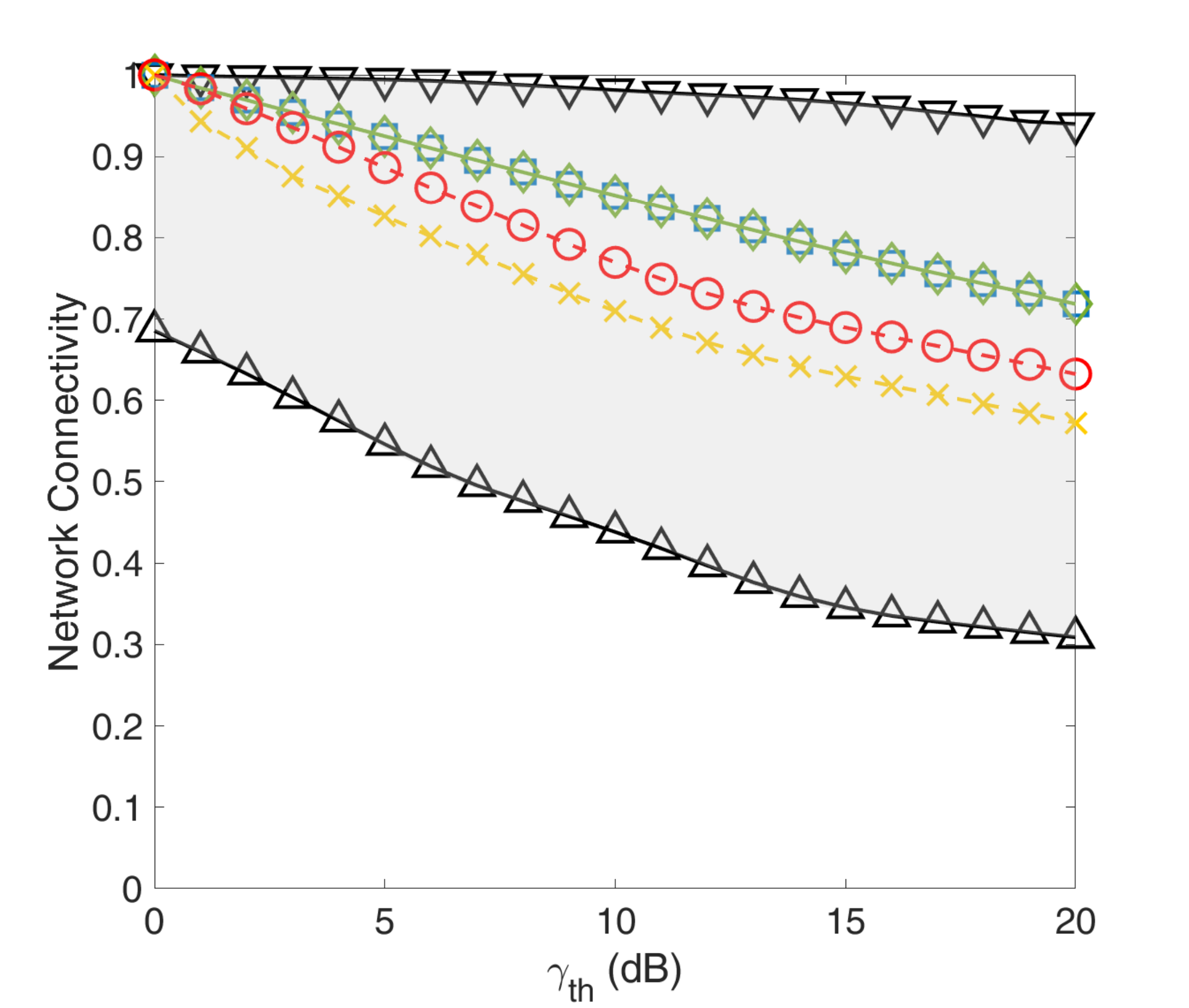}}

\caption{Network connectivity versus SNR $\gamma_{th}$ in urban roundabout: (a) medium and (b) high vehicular density for all CAVs.}
\label{fig:connect_rou}
\end{figure}

This section shows and discusses the numerical results obtained for the proactive relaying schemes, above described, in terms of network connectivity and computational complexity. 

The performance of the proactive relay of opportunity selection approaches, which leverage on the proposed LoS-map prediction, is here compared with the analytical network connectivity bounds. 
The lower bound, which accounts only for the direct links, is analytically derived based on the average adjacency matrix $\mathbf{\bar{A}}$ according to the real channel state.
The upper bound is the ideal case where \textit{all} the nodes of the V2X network can act as relays with unlimited resource capabilities. This is obtained by evaluating the second-order average adjacency matrix $\mathbf{\bar{A}}_2$, as defined in \eqref{eq:2ordAvailMatrix}. 
Once the lower and upper bound are defined, a good relaying strategy is expected to attain the upper bound, even if it is limited by relaying resource capabilities $R^{r}_c$ and it is affected by the uncertainty on the SNR link characterization.

The output of the relay selection algorithms is the adjacency matrix $\mathbf{\bar{A}_{R}}$,
whose entry $\left[\mathbf{\bar{A}_R}\right]_{(i,j)} = \left[\mathbf{\bar{A}}\right]_{(i,j)}$ if a relay has been successfully allocated, as in \eqref{eq:link_ava}. The network connectivity is obtained as in \eqref{eq:laplacian} by replacing $\mathbf{\bar{A}}$ with $\mathbf{\bar{A}_{R}}$.

The network connectivity versus the SNR threshold $\gamma_{th}$ is evaluated for the intersection in Fig.~\ref{fig:connect_int} and the roundabout in Fig.~\ref{fig:connect_rou} for two different traffic densities in the case of 100\% of CAVs.
The average network connectivity in the roundabout scenario is higher than the intersection one, since CAVs are in LoS condition with high probability. Generally, in the intersection scenario, the network connectivity upper bound is significantly below 1. This is due to the strong attenuation from blocking buildings, which severely limits the visibility of the CAVs.
The increase of vehicular density determines a more recurrent dynamic blockage of CAVs, thus leading to the degradation of the network connectivity bounds, as can be noticed by comparing scenarios with medium traffic density in Fig.~\ref{fig:Con_versus_snr_Inter_medium} and~\ref{fig:Con_versus_snr_round_medium} and high traffic density in Fig.~\ref{fig:Con_versus_snr_Inter_high} and~\ref{fig:Con_versus_snr_round_high}. The proactive relay selection schemes counteract the blockage from CAVs by guaranteeing high network connectivity that attains the upper bound. 
The proactive centralized approaches (ES and HG) always coincide and outperform the distributed ones, whose local decisions may lead to collisions and average network connectivity decrease. Centralized schemes are more robust in high traffic density compared to distributed techniques, but requiring more V2I signalling.

\begin{figure}[!t] 
\centering

\subfloat[ \label{fig:Con_versus_density_Inter_medium}]{ \includegraphics[width=0.90\columnwidth]{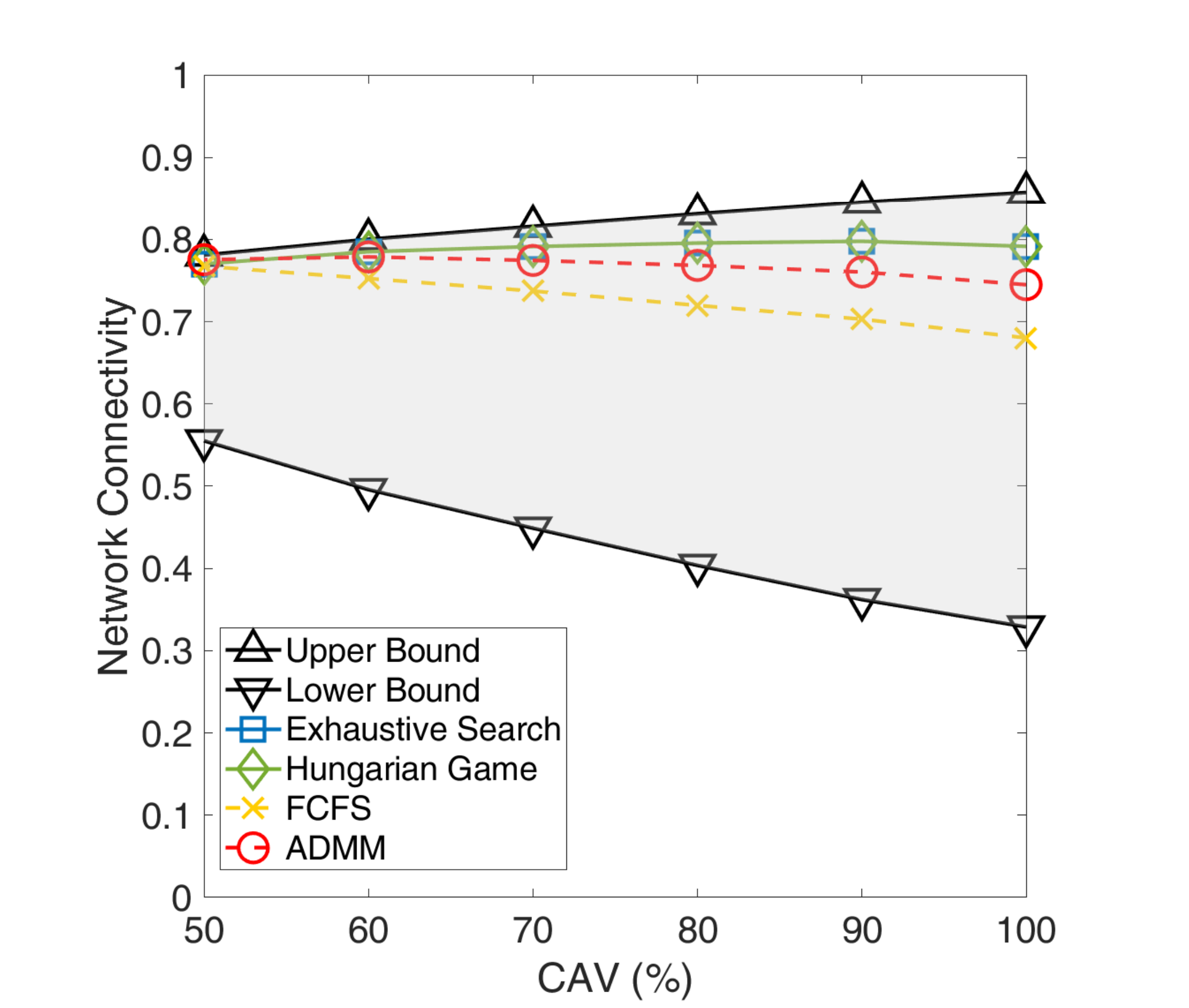}}\\
\subfloat[ \label{fig:Con_versus_density_Inter_high}]{     \includegraphics[width=0.90\columnwidth]{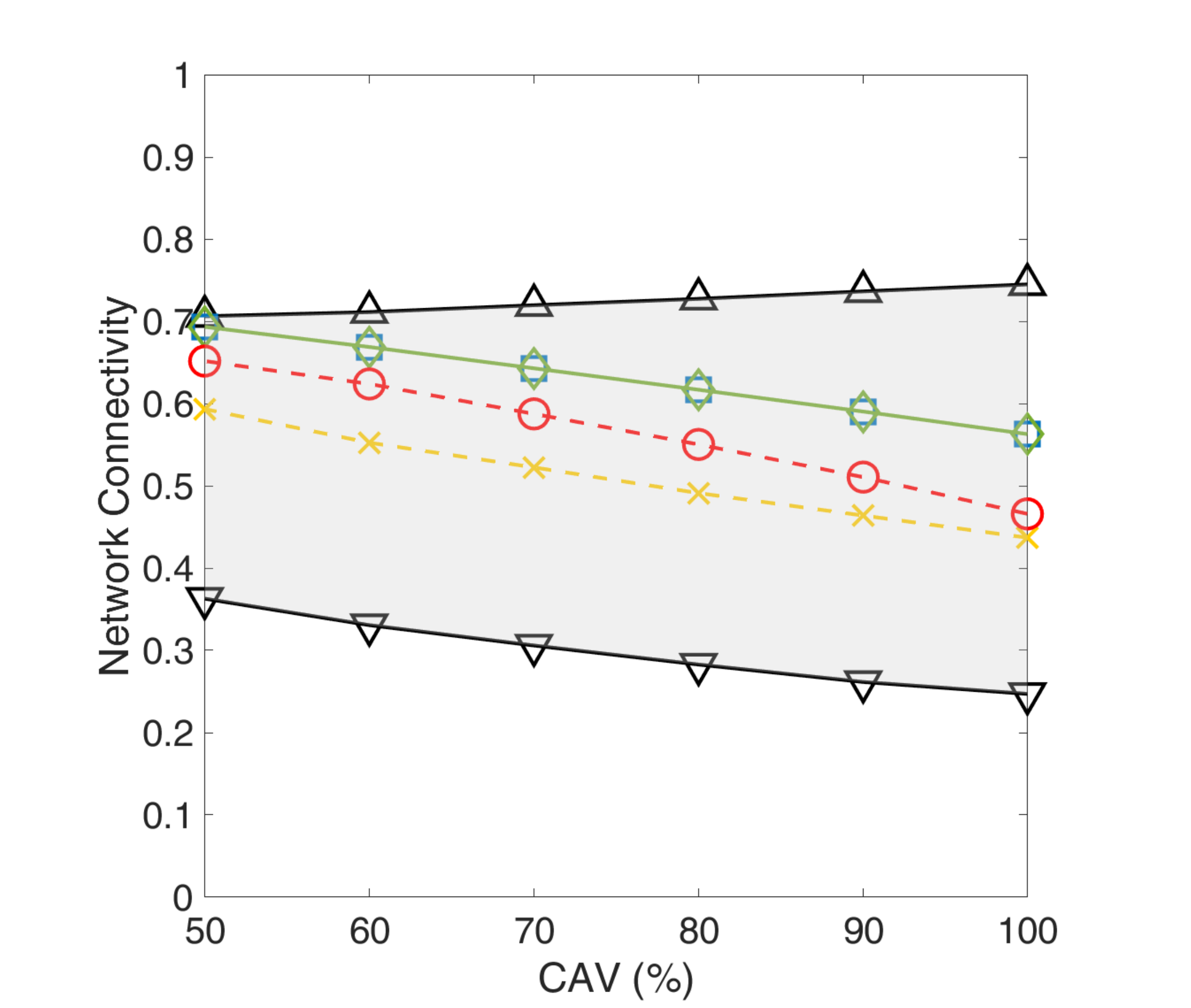}}

\caption{Network connectivity versus CAV percentage in urban intersection: (a) medium and (b) high vehicular density.}
\label{fig:connect_int_versus_density}
\end{figure}

\begin{figure}[!t] 
\centering

\subfloat[ \label{fig:Con_versus_density_round_medium}]{ \includegraphics[width=0.90\columnwidth]{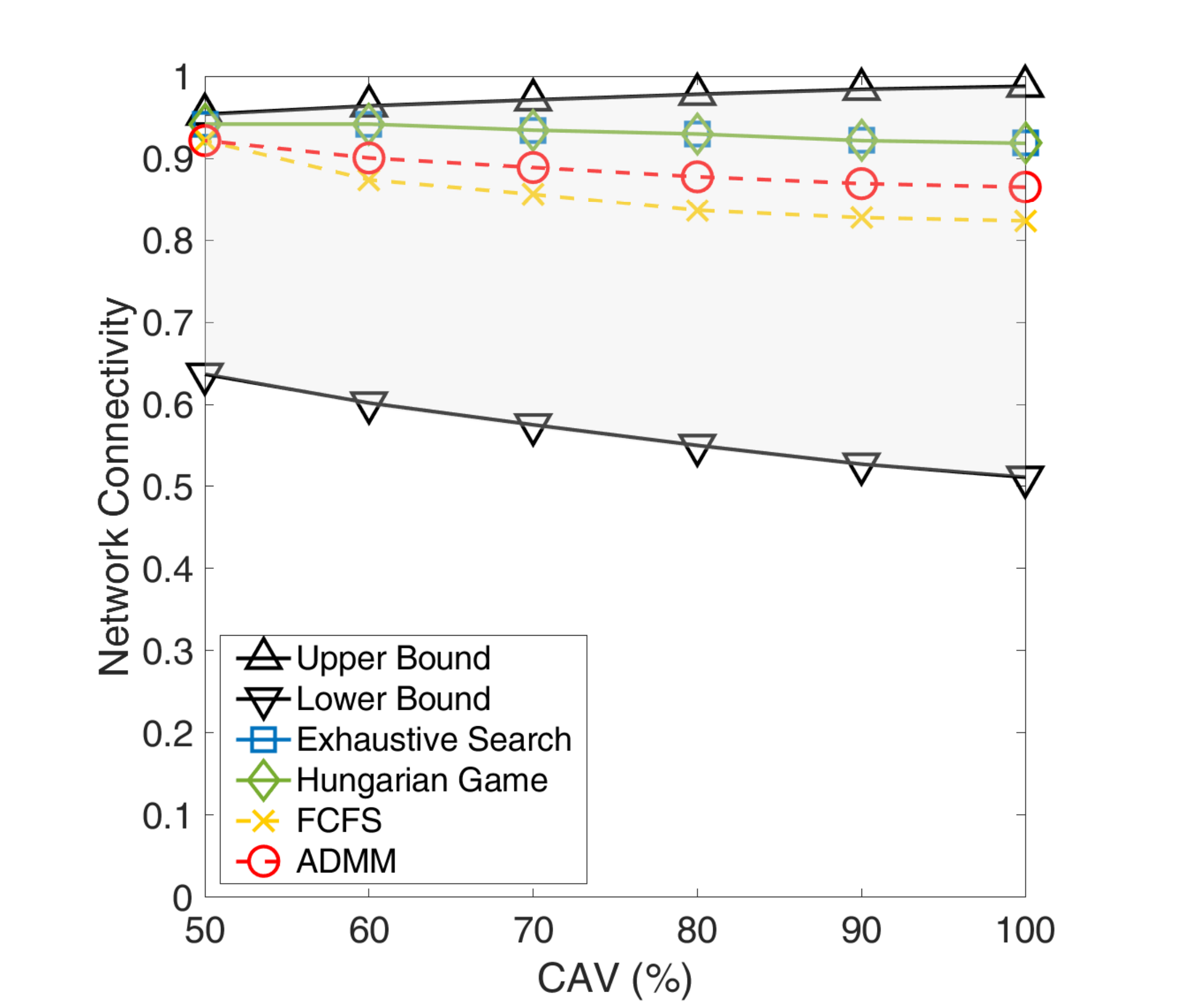}}\\
\subfloat[  \label{fig:Con_versus_density_round_high}]{     \includegraphics[width=0.90\columnwidth]{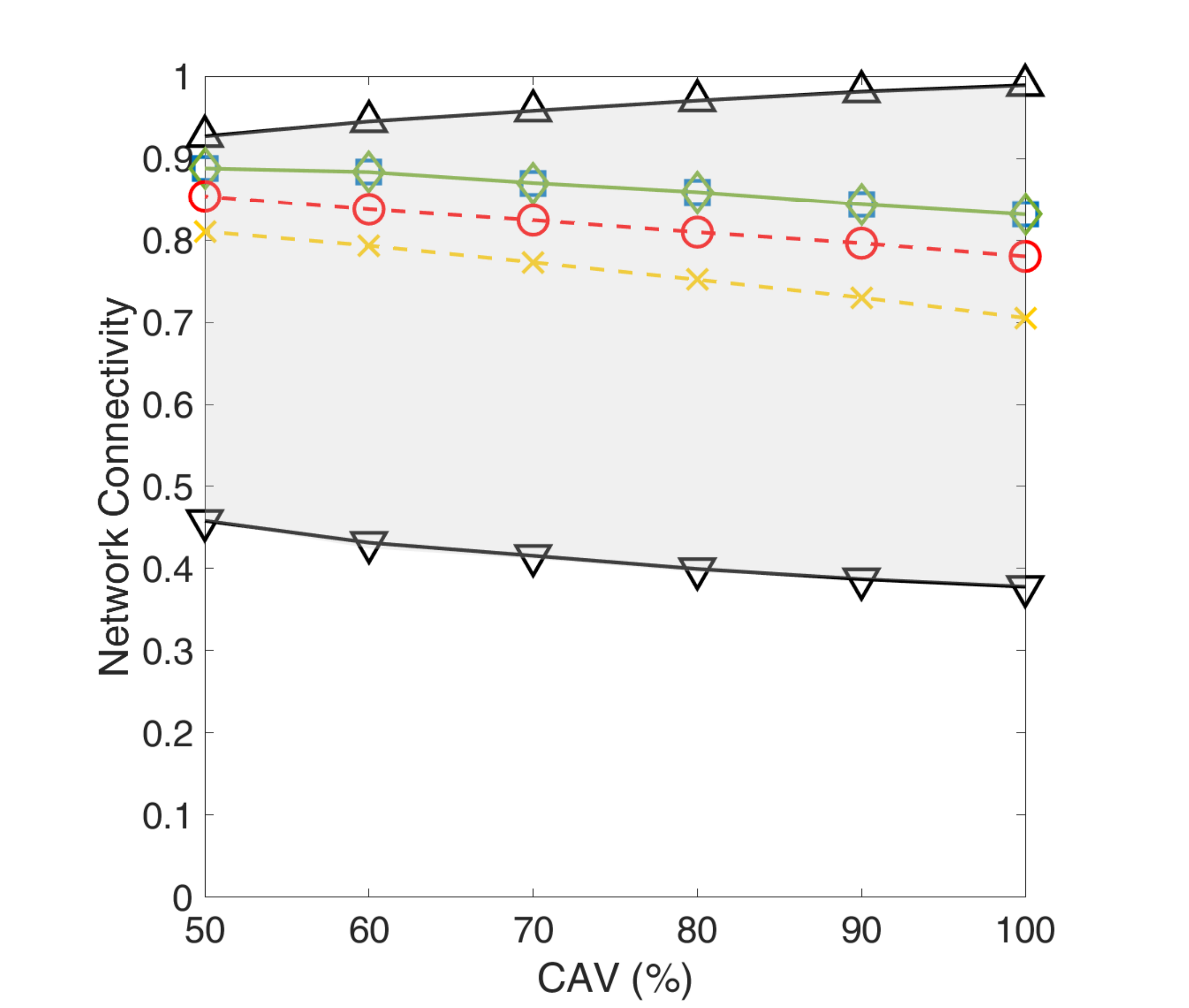}}

\caption{Network connectivity versus CAV percentage in urban roundabout: (a) medium and (b) high vehicular density.}
\label{fig:connect_round_versus_density}
\end{figure}

Figures~\ref{fig:connect_int_versus_density} and \ref{fig:connect_round_versus_density} depict the V2X network connectivity varying the density of CAVs given the SNR threshold $\gamma_{th}=10$ dB. 
As the percentage of CAVs increases, the lower bound decreases due to the high impact of recurring dynamic blockage. By contrast, the upper bound improves due to the high relaying resource capabilities, leading to an higher probability of connecting two nodes with relayed links that satisfy the SNR requirements. However, the performance of the relay selection strategies decreases as the CAVs percentage increases. This behavior is induced by the limited relay resources, suggesting that future CAVs need to support a higher communication overhead and have to increase resources for opportunistic relaying operations. 
The effect of the random blockage due to nCAVs is observed at medium and high traffic densities in Figs.~\ref{fig:Con_versus_density_Inter_medium}, \ref{fig:Con_versus_density_round_medium} and  Figs.~\ref{fig:Con_versus_density_Inter_high}, \ref{fig:Con_versus_density_round_high} for a fixed CAV percentage. The proposed proactive relay selection strategies prove to counteract the effect of the probabilistic blockage of nCAVs as they achieve an average network connectivity comparable to the upper bound, when the percentage of CAVs is around 50-70\%. 

The proactive distributed relay selection schemes are preferred in case of low percentage of CAVs and low $\gamma_{th}$ as they achieve stable network connectivity with a limited computational complexity, as described in Table~\ref{tab:complexity}. The centralized approaches are suggested in case of high CAVs density and high SNR threshold since they achieve high connectivity and they are more robust in counteracting dynamic blockage, but they require an higher computational cost.

Results here highlight the strong need to develop proactive robust and resilient relaying techniques that optimizes the network connectivity during the communication interval. Our work provides an innovative relaying framework to guarantee the SNR QoS requirements over the communication interval that maximizes the average probability of connecting nodes with reliable links.

\section{Conclusion} \label{sec:conclusions}

Millimeter-wave (mmW) multiple-input multiple-output (MIMO) technology, using beamforming, is expected to support enhanced vehicle-to-everything (V2X) services. Link quality degrades severely at high frequencies, considering that the mobility of vehicles causes frequent blockage of the line-of-sight (LoS). 
Traditional relaying techniques implement a reactive scheme to link failure and leverage only with instantaneous information, which impedes efficient relay selection in highly mobile and complex networks, such as vehicular scenarios.
In this paper, a unified framework for proactive relay selection is proposed as a tool to mitigate blockage in vehicular networks. In particular, a novel architecture for proactive relay of opportunity selection in 6G V2X is investigated. Here, the reliable lower spectrum (FR1) link enables the signaling to exchange the cooperative awareness messages and periodic position information. Based on this collected information and onboard sensors, connected and autonomous vehicles (CAVs) estimate the state of nearby objects to predict the dynamic LoS-map, which describes the network links' evolution in time.
The proactive relaying schemes exploit the dynamic LoS-map to maximize the network connectivity. 
Centralized and distributed architectures are considered and compared in terms of network connectivity and complexity. Moreover, the upper and lower bounds of the network connectivity are derived analytically and are used to compare different relaying methods.
The proposed framework and the analytical tools are validated through numerical simulations in realistic scenarios and different traffic densities. 
The results suggest that LoS blockage has a severe impact on communication in 6G vehicular networks. Further, the proactive relay-based solutions are required to mitigate blockage and maximize the average probability of connecting CAVs with reliable links.

\appendices
\begin{figure*}[t!] 
    \begin{equation}
        \label{eq:P_n_teorica}
        P_{NLoSv}
        = \iint_{\boldsymbol{\Omega}} \frac{1}{\sqrt{2 \pi \text{det}\left(\boldsymbol{\tilde{\Sigma}}_p\right)}} \,\exp{\left(-\frac{1}{2}\left(\mathbf{s}_p-\mathbf{s}_{\text{p,true}}\right)^H \,\left.\boldsymbol{\tilde{\Sigma}}_p\right.^{-1}  \left(\mathbf{s}_p-\mathbf{s}_{\text{{p},true}}\right) \right)}\, d\boldsymbol{\Omega} 
    \end{equation}
\end{figure*} 
\begin{figure*}
\begin{align} \label{eq:PnLoS}
     P_{{NLoSv}}=\left(\mathcal{Q}\left(\frac{x_1 - x_{true}}{\sigma_x}\right)-\mathcal{Q}\left(\frac{x_2 - x_{true}}{\sigma_x}\right) \right) \left(\mathcal{Q}\left(\frac{y_1 - y_{true}}{\sigma_y} \right)-\mathcal{Q}\left(\frac{y_2 - y_{true}}{\sigma_y}\right) \right)
\end{align}
\end{figure*}
\begin{figure*}
\begin{align} 
\begin{split}
\label{eq:path_los_dist}
    f_{\mathrm{{SNR}}}({\gamma}) = &\underbrace{ P_{LoS} \frac{1}{\sqrt{2 \pi \sigma^2_{sh}}} \exp{\left(-\frac{\left(\gamma-A_{LoS}(d))\right)^2}{2 (\sigma_{sh}^2)}\right)}}_{\text{LoS component} f_{{LoS}}({\gamma})}  + \\ 
    &+\underbrace{P_{{NLoSv}} \frac{1}{\sqrt{2 \pi (\sigma_{v}^2+\sigma_{sh}^2)}} \exp{\left(-\frac{\left(\gamma - (A_{LoS}(d)+A_{v})\right)^2}{2 (\sigma_{v}^2+\sigma_{sh}^2)}\right)}}_{\text{NLoSv component } f_{{NLoSv}}({\gamma})} 
    \end{split}
\end{align}
\end{figure*}
\section{}
This section presents the analytical derivation of the SNR distribution. 
The absence of nCAVs and the assumption of knowing the positions of the VoI and RV, as well as the objects' (e.g., buildings, foliage, and other CAVs) footprint, make the SNR pdf solely dependent on the shadowing component. 
On the contrary, if nCAVs are part of the considered scenario determining the NLoSv condition (see Remark 2), the SNR pdf also depends on the probability that a single nCAV (or several nCAVs) block(s)the LoS. Therefore, to analytically derive the distribution of the SNR, it is necessary to obtain this probability.

\subsection{Single nCAV Blockage}
For sake of simplicity, we first consider the case of a single nCAV, whose state $\mathbf{\tilde{s}}$ at time instant $\bar{t}$ (hereinafter omitted for clarity of notation) is modeled as in Sec.~\ref{sec:sens}. To derive the blockage probability, we consider only the position information $\mathbf{\tilde{s}_p} = [\tilde{x}, \tilde{y}]$, that is
%
\begin{align}\label{eq:statPos}
    \mathbf{\tilde{s}}_p \sim \mathcal{N}\left(\mathbf{s}_{\text{{p},true}}, \boldsymbol{\tilde{\Sigma}}_p\right),
\end{align}
where $\boldsymbol{\tilde{\Sigma}}_p$ is defined in \eqref{eq:cov}. The blockage probability $P_{{NLoSv}}$ is given in \eqref{eq:P_n_teorica}, where $\boldsymbol{\Omega}$ represents the blockage area (see Fig.~\ref{fig:unc}). 
The double integral in \eqref{eq:P_n_teorica} admits a closed form solution if the blockage area $\boldsymbol{\Omega}$\,$\approx$\,$[x_1, x_2]\times[y_1, y_2]$ is approximately rectangular and the covariance matrix $\boldsymbol{\tilde{\Sigma}}_p$\,=\,$diag\left(\sigma_x^2, \sigma_y^2\right)$ is diagonal. In this case, the probability of blockage $P_{{NLoSv}}$ is obtained in \eqref{eq:PnLoS}, where $\mathcal{Q}(\cdot)$ is the Q-function. 
The LoS probability is obtained by $P_{{LoS}}$\,=\,$1-P_{{NLoSv}}$.
The SNR distribution $f_{\mathrm{{SNR}}}(\gamma)$ is computed in \eqref{eq:path_los_dist} as a mixture of two normal distributions, which represent the two concurring propagation conditions (LoS and NLoSv). An example of \eqref{eq:path_los_dist} is depicted in Fig.~\ref{fig:sim_teo}.
To derive the service probability in \eqref{eq:servProb}, we evaluate the cdf of the SNR as
\begin{align} \label{eq:sum_of_cdf}
    F_{\text{SNR}}(\gamma_{\text{th}})= P_{{LoS}}   F_{{LoS}}(\gamma_{\text{th}}) +  P_{{NLoSv}}  F_{{NLoSv}}(\gamma_{\text{th}})
\end{align}
where $F_{{LoS}}(\gamma_{\text{th}})$ and $F_{{NoLv}}(\gamma_{\text{th}})$ are the cdfs of LoS and NLoS conditions, respectively. The cdf of a normal distribution $ x \sim \mathcal{N}(\mu, \sigma^2)$ can be computed as
\begin{align} \label{eq:cfd_sol}
    F_X(x)= 1-\mathcal{Q}(x)
\end{align}

\subsection{Multiple nCAVs Blockage}
We generalize the blockage probability derivation to the case of multiple blockers.
Assuming the presence of $N_b$ nCAVs, whose state position is defined in \eqref{eq:statPos}, the SNR distribution depends on the shadowing component and on the effective number $k_b$ of nCAVs simultaneously blocking the LoS, with 1 $\leq$ $k_b$ $\leq N_b$. This can be undertaken as a combinatorial problem. Indeed, to derive the probability $P_{NLoSv}^{(k_b)}$ that $k_b$ out of $N_b$ nCAVs are blocking the LoS, evaluation of all possible combinations is required.
The total number of combinations is given by the binomial coefficient
\begin{align}
    N{(k_b)} = \binom{N_b}{k_b}  = \frac{N_b!}{(N_b-k_b)! k_b!}.
\end{align}

We refer to the single $n$th $k_b$-tuple as $\mathbf{b}_n$, with $1 \leq n \leq N{(k_b)}$. The NLoSv probability due to $k_b$ nCAVs effective blockers is
\begin{align} \label{eq:P_nlosv_kb}
P_{NLoSv}^{(k_b)}\hspace{-0.1cm}=\hspace{-0.1cm}\sum_{n = 1}^{ N{(k_b)}}\hspace{-0.1cm}\left( \prod_{i \in \mathbf{b}_n} P_{NLoSv}^{(i)} \hspace{-0.2cm}\prod_{\substack{j \not\in \mathbf{b}_n \\ j \in [1, N_b]}} \hspace{-0.4cm}\left(1- P_{NLoSv}^{(j)}\right) \hspace{-0.1cm}\right ).
\end{align}
For example, if we consider $N_b$\,=\,$3$ nCAVs with index $i$\,=\,$\{1,2,3\}$, the probability that $k_b$ = $2$ of them block the LoS according to \eqref{eq:P_nlosv_kb} is 
\begin{align} 
\begin{split}
    P_{NLoSv}^{(2)} &= p^{(1)} p^{(2)} (1-p^{(3)}) + \\ 
    &\hspace{-0.5cm}+ p^{(1)} p^{(3)} (1-p^{(2)}) + p^{(2)} p^{(3)} (1-p^{(1)}) 
\end{split}
\end{align}
where $p^{(i)}$ is the probability that the $i$th nCAV is blocking the LoS.\\
Finally, the SNR distribution $f_{\mathrm{{SNR}}}({\gamma})$ is a mixture of Gaussians, computed as
\begin{align} \label{eq:f_snr_multiple}
    f_{\text{SNR}}(\gamma) = P_{LoS}  f_{LoS}(\gamma) + \sum_{k_b = 1}^{ N_{b}} P_{NLoSv}^{(k_b)}   f_{NLoSv}^{(k_b)}(\gamma)
\end{align}
where $f_{NLoSv}^{(k_b)}(\gamma)$ is the SNR pdf in case of $k_b$ blocking vehicles and $P_{LoS}$\,=\,$1 - \sum_{k_b = 1}^{ N_{b}} P_{NLoSv}^{(k_b)}$. The mean $A^{(k_b)}_{v}$ and the variance $\sigma^{2\,(k_b)}_{v}$ of the extra attenuation depend on $k_b$ according to the 3GPP recommendations \cite{14rel}.
Similarly, the cdf of the SNR in the multiple nCAVs blockage is obtained as
\begin{align} \label{eq:F_snr_multiple}
    F_{\text{SNR}}(\gamma_{\text{th}})&= P_{{LoS}}   F_{{ L}}(\gamma_{\text{th}}) + \sum_{k_b = 1}^{ N_{b}} P_{{NLoSv}}^{(k_b)} F_{{NLoSv}}^{(k_b)}(\gamma_{\text{th}})
\end{align}
This result is used to determine the service probability in \eqref{eq:servProb}.

\section*{Acknowledgment}

This research was carried out in the framework of the Huawei-Politecnico di Milano Joint Research Lab. The Authors want to acknowledge the Huawei Milan Research Centre.

\ifCLASSOPTIONcaptionsoff
  \newpage
\fi
\footnotesize
\bibliographystyle{IEEEtran}
\bibliography{biblio.bib}

\newpage
\begin{IEEEbiography}[{\includegraphics[width=1in,height=1.25in,clip,keepaspectratio]{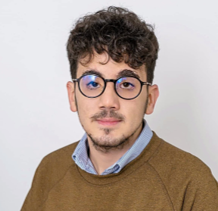}}]{Francesco Linsalata} received B.Sc. and M.Sc. degrees cum laude in Telecommunication engineering from Politecnico di Milano, Milan, Italy, in 2017 and 2019, respectively. He is PhD student at the Dipartimento di Elettronica, Informazione e Bioingegneria, Politecnico di Milano.  His main research interests focus on V2X communications and waveforms design for B5G wireless networks. He was the co-recipient of the best-paper award and recipient of the best student paper award at BalkanCom'19. 
\end{IEEEbiography}

\begin{IEEEbiography}[{\includegraphics[width=1in,height=1.25in,clip,keepaspectratio]{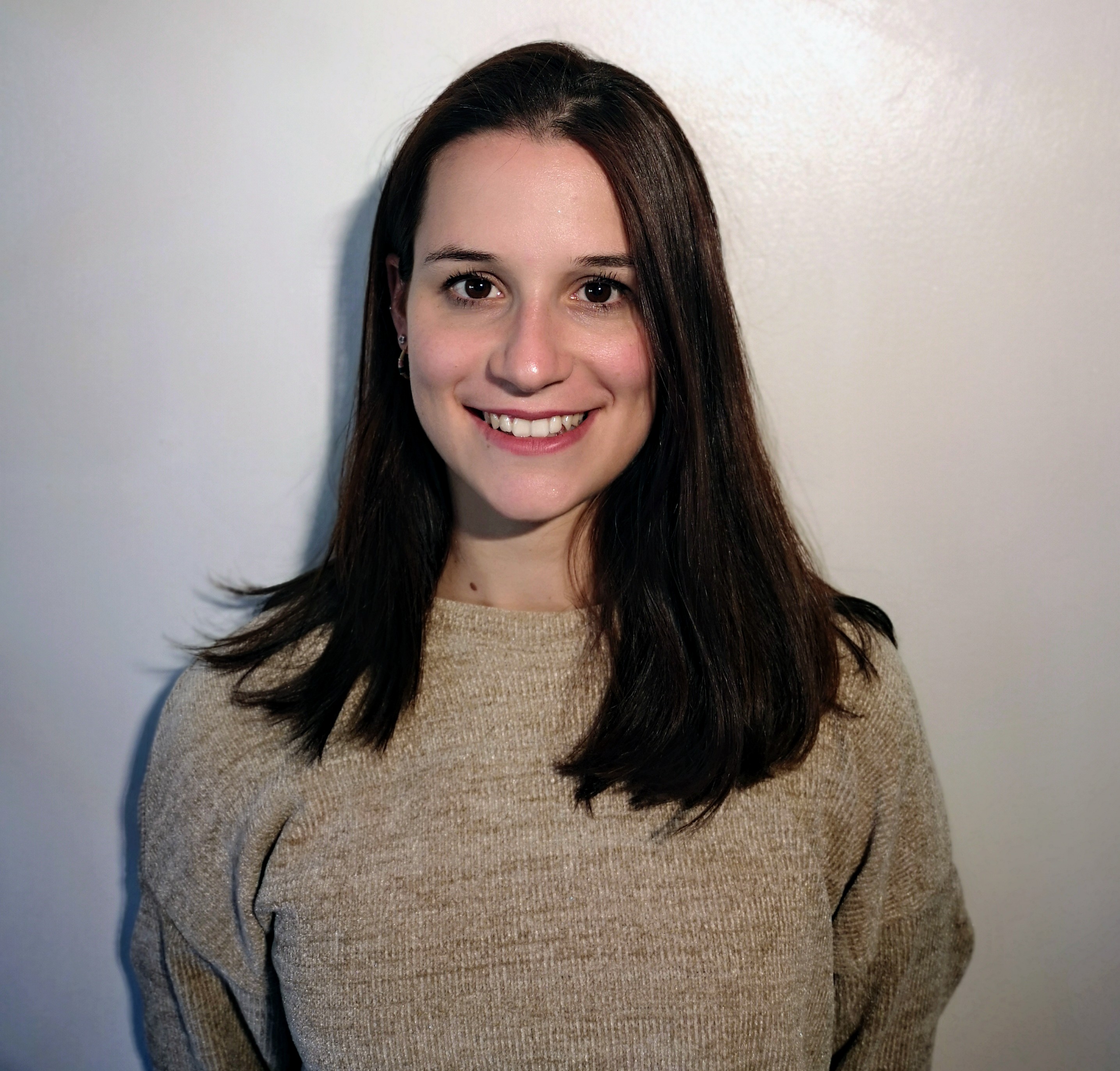}}]{Silvia Mura} received the M.Sc. degree in Telecommunication engineering from Politecnico di Milano, Milan, Italy, in 2020. She is a PhD student at the Dipartimento di Elettronica, Informazione e Bioingegneria, Politecnico di Milano.  Her main research interests focus on signal processing in V2X communications. 
\end{IEEEbiography}

\begin{IEEEbiography}[{\includegraphics[width=1in,height=1.25in,clip,keepaspectratio]{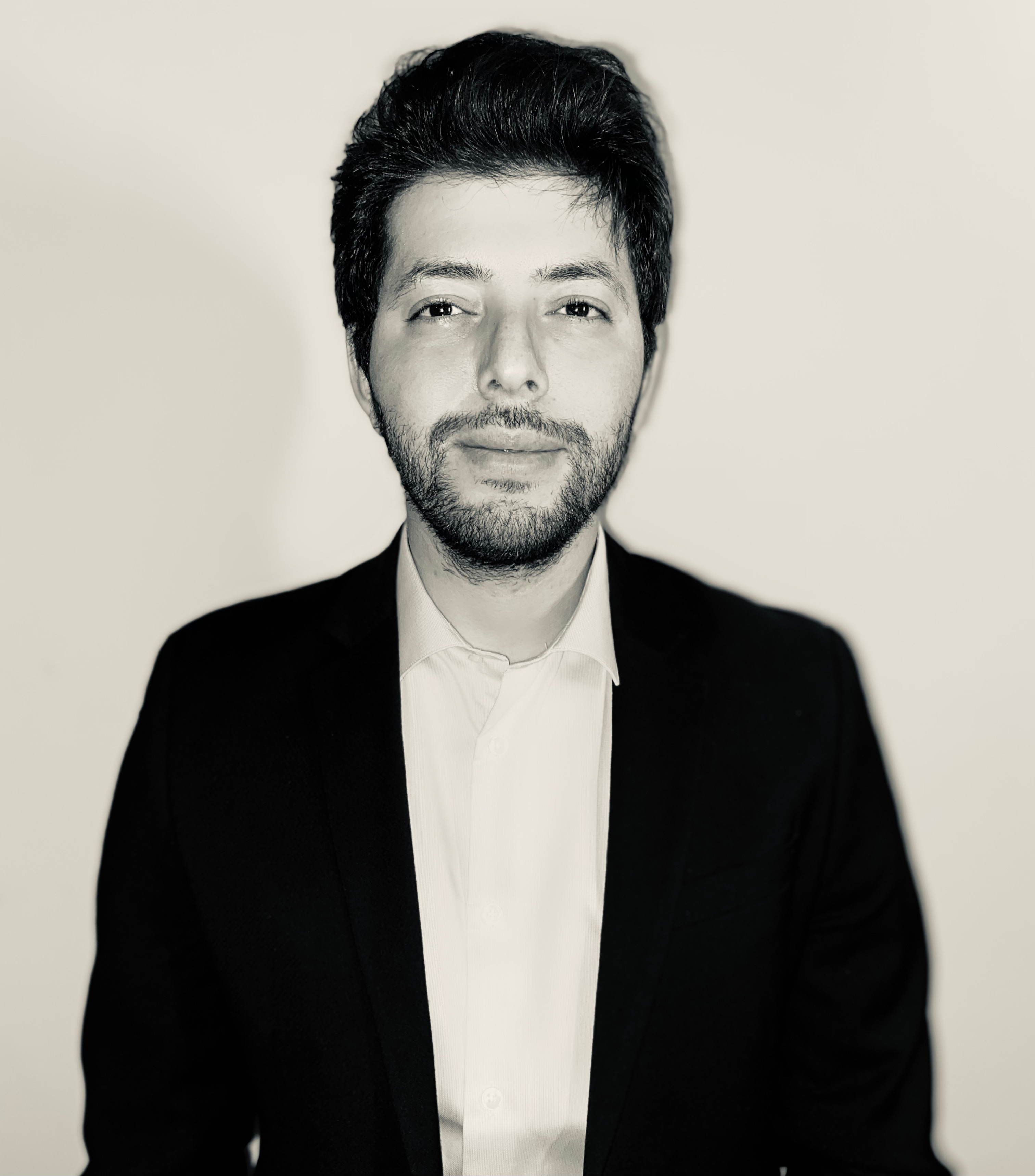}}] {Marouan Mizmizi} is a post-doctoral research fellow at Dipartimento di Elettronica, Informazione e Bioingegneria, Politecnico di Milano, Italy, in the framework of Huawei-Polimi Joint Research Lab. He received the B.Sc. degree (2012), M.Sc. degree (2015) in Telecommunication Engineering and the Ph.D. (2019) in Information Technology from Politecnico di Milano. His research interests comprise advanced signal processing techniques for V2X communication systems, in particular for MIMO channel estimation in mmWave and sub-THz frequencies and Relay selection and scheduling in high mobility scenarios. He was the recipient of the Best Paper Award from the 29th International Conference on Advanced Information Systems Engineering, 2017.
\end{IEEEbiography}

\begin{IEEEbiography}[{\includegraphics[width=1in,height=1.25in,clip,keepaspectratio]{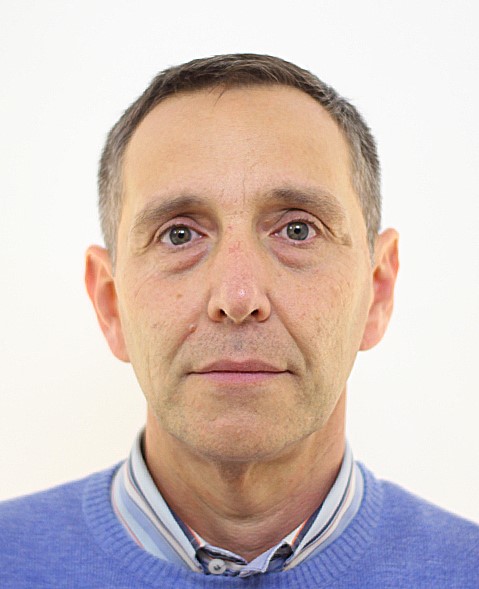}}] {Maurizio Magarini} (M'04) received the M.Sc. and Ph.D. degrees in electronic engineering from the Politecnico di Milano, Milan, Italy, in 1994 and 1999, respectively. In 1994, he was granted the TELECOM Italia scholarship award for his M.Sc. Thesis. He worked as a Research Associate in the Dipartimento di Elettronica, Informazione e Bioingegneria at the Politecnico di Milano from 1999 to 2001. From 2001 to 2018, he was an Assistant Professor in Politecnico di Milano where, since June 2018, he has been an Associate Professor. From August 2008 to January 2009 he spent a sabbatical leave at Bell Labs, Alcatel-Lucent, Holmdel, NJ. His research interests are in the broad area of communication and information theory. Topics include synchronization, channel estimation, equalization and coding applied to wireless and optical communication systems. His most recent research activities have focused on molecular communications, massive MIMO, study of waveforms for 5G cellular systems, vehicular communications, wireless sensor networks for mission critical applications, and wireless networks using unmanned aerial  vehicles and high-altitude platforms. He has authored and coauthored more than 120 journal and conference papers. He was the co-recipient of two best-paper awards. He is an Associate Editor of IEEE Access, IET Electronics Letters, and Nano Communication Networks (Elsevier). He has been involved in several European and National research projects.
\end{IEEEbiography}

\begin{IEEEbiography}[{\includegraphics[width=1in,height=1.25in,clip,keepaspectratio]{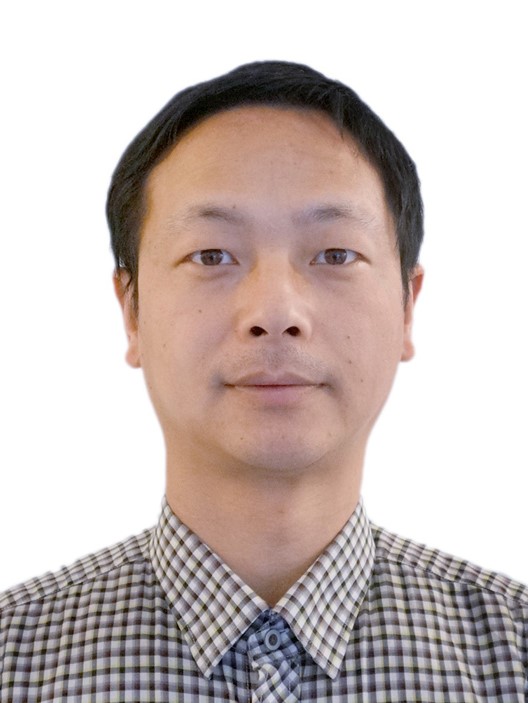}}]
{Peng Wang} received the Ph.D. degree in electronic engineering from the City University of Hong Kong, Hong Kong SAR, in 2010. He was a Research Fellow with the City University of Hong Kong from 2010 to 2012, and a Research Fellow with the University of Sydney, Australia, from 2012 to 2015. Since 2015, he has been with Huawei Technologies, Sweden, where he is currently a Senior Research Engineer. He has authored or co-authored over 60 peer-reviewed research papers in the leading international journals and conferences. His current research interests include 5G and beyond standardization, millimeter-wave communication, MIMO techniques, information theory and iterative multiuser detection. He received the Best Paper Award at the 2014 IEEE International Conference on Communications. 
\end{IEEEbiography}

\begin{IEEEbiography}[{\includegraphics[width=1in,height=1.25in,clip,keepaspectratio]{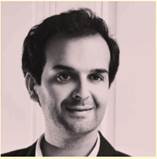}}]{Majid Nasiri Khormuji}  received the Ph.D. degree in telecommunications from KTH—Royal Institute of Technology, Stockholm, Sweden in 2011. He held a visiting position at Stanford University, Stanford, CA in 2011. From 2011 to 2013, he was a Postdoctoral Research Fellow at the School of Electrical Engineering and the ACCESS Linnaeus Center at KTH. Since 2013, he is a Senior Researcher at Huawei Technologies Sweden AB in Stockholm, wherein he is conducting fundamental research for future wireless communication systems. Dr. Nasiri Khormuji has worked on problems in the area of network information theory, coding and transmission for wireless communications and statistical signal processing. He is the author of more than 50 journal and conference papers, some of which received best paper awards, and three chapters in books. He is also the inventor of numerous patent applications in a wide range of technologies spanning Multiple-Access, Modulation, Coding, MIMO, Massive MIMO and SWIPT. He has also served on the technical program committees for IEEE sponsored conferences including ICC, Globecom and WCNC.
\end{IEEEbiography}

\begin{IEEEbiography}[{\includegraphics[width=1in,height=1.25in,clip,keepaspectratio]{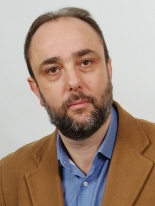}}] {Alberto Perotti} is a Principal Research Engineer at Huawei Technologies, where he is involved in wireless networks’ PHY layer research and standardization.
He received his M.Sc. and Ph.D. degrees in telecommunications from Politecnico di Torino, Italy, in 1999 and 2003,respectively.
Previously, he carried out research on mobile wireless and satellite communications at Politecnico di Torino.
His interests cover coding and modulation, multiple access, and software-defined radios.
\end{IEEEbiography}

\begin{IEEEbiography}[{\includegraphics[width=1in,height=1.25in,clip,keepaspectratio]{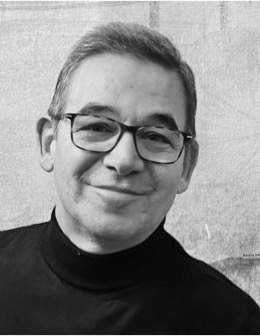}}] {Umberto Spagnolini} is Professor of Statistical Signal Processing, Director of Joint Lab Huawei-Politecnico di Milano and Huawei Industry Chair. His research in statistical signal processing covers remote sensing and communication systems with more than 300 papers on peer-reviewed journals/conferences and patents. He is author of the book Statistical Signal Processing in Engineering (J. Wiley, 2017). The specific areas of interest include mmW channel estimation and space-time processing for single/multi-user wireless communication systems, cooperative and distributed inference methods including V2X systems, mmWave communication systems, parameter estimation/tracking, focusing and wavefield interpolation for remote sensing (UWB radar and oil exploration). He was recipient/co-recipient of Best Paper Awards on geophysical signal processing methods (from EAGE), array processing (ICASSP 2006) and distributed synchronization for wireless sensor networks (SPAWC 2007, WRECOM 2007). He is technical experts of standard-essential patents and IP. He served as part of IEEE Editorial boards as well as member in technical program committees of several conferences for all the areas of interests.
\end{IEEEbiography}

\end{document}